\documentclass[journal,cspaper,compsoc]{IEEEtran}

\usepackage{graphicx}
\usepackage{balance}
\usepackage{comment}
\usepackage{cite}
\usepackage{amssymb}
\usepackage[tight,footnotesize]{subfigure}
\usepackage[active]{srcltx}
\usepackage{amsmath}
\usepackage{amsfonts}
\newcommand{\vecbold}[1]{\ensuremath{\boldsymbol{#1}}}
\newcommand{\norm}[1]{\left\lVert#1\right\rVert}
\graphicspath{{./figs/}}
\renewcommand{\baselinestretch}{0.95}
\usepackage{color,soul}
\usepackage{eurosym}
\usepackage{algorithm}
\usepackage{algorithmic}
\usepackage{multicol}
\usepackage{xcolor}
\usepackage{flushend}
\usepackage{multicol}

\begin{document}

\title{Jointly-optimized Searching and Tracking with Random Finite Sets}

\author{Savvas~Papaioannou,~Panayiotis~Kolios,~Theocharis~Theocharides,\\~Christos~G.~Panayiotou~ and ~Marios~M.~Polycarpou

\IEEEcompsocitemizethanks{\IEEEcompsocthanksitem The authors are with the KIOS Research and Innovation Centre of Excellence (KIOS CoE) and the Department of Electrical and Computer Engineering, University of Cyprus, Nicosia, 1678, Cyprus.\protect\\
E-mail:\texttt{\{papaioannou.savvas, pkolios, ttheocharides, christosp, mpolycar\}@ucy.ac.cy}}
\thanks{}}

\markboth{IEEE Transactions on Mobile Computing, vol. 19, no. 10, pp. 2374-2391, 1 Oct. 2020}%
{Papaioannou \MakeLowercase{\textit{et al.}}: Jointly-optimized Searching and Tracking with Random Finite Sets}

\IEEEcompsoctitleabstractindextext{%
\begin{abstract}
In this paper we investigate the problem of joint searching and tracking of multiple mobile targets by a group of mobile agents. The targets appear and disappear at random times inside a surveillance region and their positions are random and unknown. The agents have limited sensing range and receive noisy measurements from the targets. 
A decision and control problem arises, where the mode of operation (i.e. \textit{search} or \textit{track}) as well as the mobility control action for each agent, at each time instance, must be determined so that the collective goal of searching and tracking is achieved. We build our approach upon the theory of random finite sets (RFS) and we use Bayesian multi-object stochastic filtering to simultaneously estimate the time-varying number of targets and their states from a sequence of noisy measurements. We formulate the above problem as a non-linear binary program (NLBP) and show that it can be approximated by a genetic algorithm. Finally, to study the effectiveness and performance of the proposed approach we have conducted extensive simulation experiments. 
\end{abstract}

\begin{IEEEkeywords}
	Bayesian target tracking, Intelligent systems, Sensor control, Area coverage
\end{IEEEkeywords}}

\maketitle

\IEEEdisplaynotcompsoctitleabstractindextext
\IEEEpeerreviewmaketitle

\section*{Nomenclature}
\addcontentsline{toc}{section}{Nomenclature}
\begin{IEEEdescription}[\IEEEusemathlabelsep\IEEEsetlabelwidth{$pi_{k|k-1}(x_k|x_{k-1})$}]
	\item[$x_k \in \mathcal{X}$] Single target state vector at time $k$.
	\item[$z_k \in \mathcal{Z}$] Measurement vector at time $k$.
	\item[$s_k \in \mathcal{A}$] Single agent state vector at time $k$
	\item[$u_k \in \mathbb{U}_k$] Single agent control action at time $k$.
	\item[$X_k \in \mathcal{F}(\mathcal{X})$] Multi-target state (i.e. finite set) at time $k$.
	\item[$Z_k \in \mathcal{F}(\mathcal{Z})$] Measurement set (i.e. finite set) at time $k$.
	\item[$\pi_{k|k-1}(x_k|x_{k-1})$] Transitional density of $x_k$ based on time $k-1$ given $x_{k-1}$.
	\item[$g_k(z_k|x_k,s_k)$] Measurement likelihood function of $z_k$ conditioned on $x_k$ and $s_k$.
	\item[$p_D(x_k,s_k)$] Single agent sensing model i.e. probability of detection.
	\item[$p_{k|k-1}(x_k|z_{1:k-1})$] Predictive distribution of $x_k$ based on time $k-1$ given $z_{1:k-1}$.
	\item[$p_{k}(x_k|z_{1:k})$] Posterior distribution of $x_k$ at time $k$ given $z_{1:k}$.
	\item[$f_{k|k-1}(X_k|Z_{1:k-1})$] Multi-object predictive distribution of $X_k$ based on time $k-1$ given $Z_{1:k-1}$.
	\item[$f_{k}(X_k|Z_{1:k})$] Multi-object posterior distribution of $X_k$ given $Z_{1:k}$.
	\item[$r_{k|k-1}$] Single target probability of existence at time $k$ based on time $k-1$ (prior).
	\item[$r_{k}$] Single target probability of existence at time $k$ (posterior).
	\item[$v_{k|k-1}$] Maximum number of hypothesized targets at time $k$ based on time $k-1$.
	\item[$v_{k}$] Maximum number of hypothesized targets at time $k$ (posterior).
	\item[$Z_{k|k-1}$] Predicted ideal measurement set (PIMS) for time $k$ based on time $k-1$.
	\item[$\hat{X}_{k|k-1}$] Estimation of $X_k$ based on time $k-1$.
	\item[$\hat{X}_{k}$] Estimation of $X_k$ based on time $k$.
	\item[$\hat{n}_{k|k-1}$] Estimation of the true number of targets $n_k$ based on time $k-1$.
	\item[$\hat{n}_{k}$] Estimation of the true number of targets $n_k$ based on time $k$.
	\item[$\sigma_{k}$] Variance of the estimated number of targets $\hat{n}_{k}$
	\item[$\tilde{\sigma}_k$] Normalized variance of the estimated number of targets $\hat{n}_{k}$
	\item[$S = \{1,2,...,|S|\}$] The finite set of all agents for searching and tracking.
	\item[$\xi_\text{search}$] Multi-agent searching objective function.
	\item[$\xi_\text{track}$] Multi-agent tracking objective function.
\end{IEEEdescription}

\section{Introduction}
The main objective of a search and rescue mission (e.g. ground search and rescue, air-sea rescue, etc) is to search for and provide aid to people who are in imminent danger as efficiently and safely as possible. When disasters happen e.g. earthquakes, marine disasters and aircraft accidents, the search and rescue (SAR) missions are of critical importance for finding survivors and saving lives. SAR missions however could be very dangerous and expensive (e.g. the search for the Malaysian Airlines Flight 370 that disappeared on 8 March 2014 had costed a total of more than US \$150 million).

On the other hand, the miniaturization and cost reduction of electronic components and the recent advances in robotics and specifically in small unmanned aerial vehicles (UAVs) have spurred an unprecedented interest on intelligent mobile agents for group missions. Motivated by this, we believe that a team of autonomous mobile agents could become an important aid in many search and rescue missions by improving the efficiency while at the same time reducing the need to place the rescuers in dangerous situations. Take for instance a maritime disaster where a team of drones could be deployed to search for and track survivors until further assistance is available. 

In this work we are interested in the task of searching for and tracking multiple survivors or targets of interest in SAR missions with a group of autonomous mobile agents. This is a challenging problem since a) the number of survivors that needs to be tracked at a time instance is random, unknown and thus needs to be estimated, b) the agents need to estimate the location of the survivors through noisy measurements and c) the mobile agents exhibit a limited sensing range thus there is a need for efficient searching. Finally, the problem becomes even more challenging due to the tradeoff which arises on how many resources to allocate in each task i.e. \textit{searching} or \textit{tracking} at each time instance so that the joint search and track (SAT) objective is achieved as best as possible.

To address the above challenges, we propose a unified probabilistic approach to jointly tackle the problem of searching and tracking of multiple moving targets by a team of mobile agents. The proposed approach, refereed to hereafter as JoSAT (Jointly-optimized Searching and Tracking) is formulated in the framework of recursive Bayesian multi-object stochastic estimation using random finite sets (RFS) \cite{Mahler2004_2,Mahler2013} which allows us a) to simultaneously estimate the time-varying number of targets and their states from a sequence of noisy measurements and b) avoids the problem of data-association in multi-target tracking. The contributions of this paper are:
\begin{itemize}
	\item Provides a multi-agent probabilistic framework for jointly searching and tracking multiple targets inside a given surveillance area. We utilize the theory of random finite sets (RFS) in a multi-agent framework to accurately capture the inherent uncertainty present in many search-and-track (SAT) missions.
	\item Develops a novel decision (i.e. the agents can switch between \textit{search} and \textit{track} mode during the mission) and control (i.e. control the movement of all agents) algorithm which takes into account the stochasticity of the system in order to tackle the joint objective of searching and tracking.
	\item Formulates the decision and control problem as a non-linear binary optimization program which is then solved using a genetic algorithm.
\end{itemize}

The rest of the paper is organized as follows. Section \ref{sec:Related_Work} reviews the existing literature on searching and tracking by single and multiple agents, demonstrating in the way the contributions of this work. Section \ref{sec:Problem_Definition} outlines the problem addressed in this paper and Section \ref{sec:Sys_Arch} presents an overview of the proposed system. Section \ref{sec:Background} provides a brief overview on the framework that the proposed approach is based on namely a) random finite sets (RFS) and b) Bayesian multi-object stochastic filtering and then Section \ref{sec:Problem_Formulation} presents the details of the proposed approach. Section \ref{sec:Optimization} reformulates the problem in the context of mathematical programming, demonstrates the complexity of the resulting non-linear binary program and elaborates on how the problem can be solved in practice using a genetic algorithm. Section \ref{sec:Evaluation} conducts an extensive performance analysis and finally, Section \ref{sec:Conclusion} concludes the paper and discusses future work.

\section{Related Work}
\label{sec:Related_Work}
Coordinated teams of agents unlock significantly greater capabilities than what is possible by single-handed missions. By taking this into consideration, existing literature has looked into several challenging problems that multi-agent systems can successfully tackle.

The coverage problem is among the most cited problems, where a fleet of airborne agents need to spread across an area over the shortest time interval for situational awareness or when searching for particular targets. Research works have focused around task assignment, scheduling and path planning of the multiple agents, considering physical resource constraints such as the total number of agents, their battery levels and their communication ranges \cite{Vera2015}\cite{Khan2015}\cite{Avellar2015}\cite{Ding2017}\cite{Cassandras2017_2}\cite{Rahili2018}.

Another challenging problem is that of localization and tracking of single or multiple targets  \cite{linder2016} \cite{Papaioannou2015} \cite{Papaioannou2017}. For this problem both centralized and distributed solutions are being investigated for the envisioned tracking strategies which generally seek to minimize the tracking error for all detected targets \cite{Dutta2017} \cite{Xi2016} \cite{Meyer2015}.


Interestingly though, many of these problems are interconnected and interrelated and thus need to be looked at jointly especially in certain settings such as in search and rescue missions (SAR) where efficiency and effectiveness are of essence. The problem of SAR with single or multiple robots has attracted a wide interest in the recent years with the research community proposing methods from multi-robot cooperative learning \cite{Liu2016} and path-planning strategies \cite{Macwan2015} to collaborative online task-planning by heterogenous robot teams \cite{Beck2016} and autonomous multi-UAV systems for SAR missions \cite{Scherer2015}. Detailed surveys on multi-robot SAR can be found in \cite{Liu2013,Shah2004}. In this paper however, we are focusing on a particular sub-problem i.e. the problem of search and track which has also attracted a wide interest by the research community because of it importance in SAR missions. 
An interesting work in \cite{Bourgault2006} develops a probabilistic approach for the SAT problem but only for the single-agent single-target case. Similarly, a more recent work \cite{Liu2017} investigates the SAT problem for the single-agent single-target scenario, and proposes a model-predictive control (MPC) framework for the path-planning problem.
The work \cite{Sinha2005} considers the cooperative management of groups of UAVs as an optimization problem of the information obtained through multi-target tracking. Compared to our work however, the work in \cite{Sinha2005}, is mainly focused on improving the tracking accuracy without including the search objective.
The works in \cite{Hoffmann2006,Hoffmann2010} develop an information-theoretic multi-agent SAT approach which is used to search for a single stationary target by minimizing the entropy of the target distribution at each time step. 
Moreover, \cite{Pitre2012} looked at the problem of route planning for multi-agent SAT missions and proposed a Fisher information based approach to route planning. However, in \cite{Pitre2012} the problems of false alarms and data association are not considered. Similarly, the work in \cite{Peterson2017} presents a receding horizon controller (RHC) approach to jointly drive a group of UAVs to SAT missions. In contrast, in the proposed approach the data-association problem can be avoided altogether. Notably, the classical approaches to multi-target tracking require to first solve the data association problem and then estimate the states of the targets. RFS-based approaches can estimate the states of the targets directly without requiring to first solve the data-association problem \cite{Mahler2007book}. In addition, existing techniques make weak assumptions regarding the number of targets, the target dynamics and the measurement models.

Complementary to the aforementioned studies, the proposed system advances the state-of-the-art, by 
considering a multi-agent approach where the agents can dynamically switch their mode of operation between searching and tracking during the mission in order to jointly optimize the coupled objective of searching and multi-target tracking and by developing a unified probabilistic search-and-track framework based on random finite sets which avoids the problem of data-association and accounts for the uncertainties in the number of targets, target dynamics and  measurement models. Finally, this work
	 derives both optimal and heuristic solutions to solve the resulting joint searching and tracking problem.

\section{Problem Definition}
\label{sec:Problem_Definition}

In this work we assume that the agents can operate in either \textit{search} or \textit{track} mode and that they can switch modes dynamically during the mission in order to optimize the joint search and track objective. 
The problem to solve can be stated as follows: 
\textit{At each time step a fixed number of agents must find their mode of operation and the control actions which result in the best tradeoff between maximizing the search coverage and tracking all targets found in the surveillance area}.

\color{black} Throughout this paper we consider the following modeling assumptions.

\subsection{Single Target Dynamics}\label{ssec:single_target_dynamics}
The single target state vector $x_k \in \mathcal{X}, k \in \mathbb{N}$  evolve in time according to the following equation:
\begin{equation}\label{eq:single_dynamics}
	x_k = \zeta_{k}(x_{k-1}) + \text{v}_{k}
\end{equation} 

\noindent where the $\mathcal{X} \subseteq R^{n_x}$ denotes the state space and the known function $\zeta_{k} : \mathbb{R}^{n_x} \rightarrow \mathbb{R}^{n_x}$ models the dynamical behavior of the target. Eqn. (\ref{eq:single_dynamics}) describes the evolution of the state vector as a first order Markov process with transitional density $\pi_{k|k-1}(x_k|x_{k-1}) = p_\text{v}(x_k - \zeta_{k}(x_{k-1}) )$. The random process $\text{v}_{k} \in \mathbb{R}^{n_x}$, which is referred to as the process noise, is IID according to the probability density function $p_\text{v}(.)$. The role of the process noise is to model random disturbances in the evolution of the state. 
Without loss of generality, in this paper we assume that the state vector $x_k \in \mathcal{X} \subseteq \mathbb{R}^4$ is composed of position and acceleration components i.e. $x_k = [p_x,\dot{p}_x,p_y,\dot{p}_y]^\top$ where $(p_x,p_y)$ give the 2D position of the target in Cartesian coordinates and $(\dot{p}_x,\dot{p}_y)$ are the velocities of the target in the $x$ and $y$ direction respectively.

Suppose now that multiple targets $x_k^i$ exists inside the surveillance region where we assume that the targets are independent of each other. The targets can spawn from anywhere in the state space $\mathcal{X}$ and target births and deaths occur at random times. This means that at each time $k$, there exist $n_k$ target states $x^1_k, x^2_k,...,x^{n_k}_k$, each taking values in the state space $\mathcal{X}$ where both the number of targets $n_k$ and their individual states $x_k^i, \forall i$ are random and time-varying. The group of target states at time $k$ can now be modeled as the set:
\begin{equation}
    X_k = \{x^1_k, x^2_k,...,x^{n_k}_k\} ~ \in \mathcal{F}(\mathcal{X})
\end{equation}
where $\mathcal{F}(\mathcal{X})$ denotes the space of all finite subsets of $\mathcal{X}$. Note here that $n_k$ is the true but unknown number of targets at time $k$ which needs to be estimated along with the target states $x_k^i, \forall i$. In Sec. \ref{sec:Background}-\ref{sec:Problem_Formulation} it will become evident that $X_k$ is actually a random finite set (RFS). We will refer to $X_k$ as the \textit{multi-target state}.

\subsection{Single Agent Sensing Model} \label{ssec:measM}
The ability of an agent to sense its 2D environment is modeled by the function $p_D(x_k,s_k)$ that measures the probability that a target with state $x_k$ at time $k$ is detected by an agent with state (i.e. 2D position coordinates) $s_k =[s_x,s_y]^\top_k \in \mathcal{A} \subseteq R^2$. More specifically the detection probability of a target with state $x_k$ at position $\textit{p}_k = Hx_k$ (where $H$ is a matrix that extracts the $xy$-coordinates of a target from its state vector) from an agent with state $s_k$, is given by:
\begin{equation}\label{eq:sensing_model}
 p_D(x_k,s_k) = 
  \begin{cases} 
   p^{\max}_D & \text{if } d_k < R_0 \\
   \text{max}\{0, p^{\max}_D - \eta(d_k-R_0) \} & \text{if } d_k \ge  R_0
  \end{cases}
\end{equation}
where $d_k=\norm{H x_k-s_k}_2$ denotes the Euclidean distance between the agent and the target, $p^{\max}_D$ is the detection probability for targets that reside within $R_0$ distance (i.e. the agents primary radius) from the agent's position and finally $\eta$ captures the reduced effectiveness of the agent to detect distant targets. 

When an agent detects a single target it receives a measurement vector $z_k \in \mathcal{Z}$ which is related to the target state as follows:
\begin{equation}\label{eq:meas_model}
	z_k = h_{k}(x_k,s_k) + w_k
\end{equation}
where $\mathcal{Z} \subseteq \mathbb{R}^{n_z}$ denotes the measurement space and the function $h_{k} : \mathbb{R}^{n_x} \rightarrow \mathbb{R}^{n_z}$ projects the state vector to the measurement space. The random process $w_{k} \in \mathbb{R}^{n_z}$ is IID, independent of $\text{v}_{k}$ and distributed according to $p_w(.)$. The probability density of measurement $z_k$ for a target with state $x_k$ when the agent is at state $s_k$ is given by the measurement likelihood function $g_k(z_k|x_k,s_k) = p_w(z_k - h_k(x_k,s_k))$.
Without loss of generality, in this paper we assume that the measurement vector $z_k \in \mathcal{Z}$ consists of range and bearing measurements. 

That said an agent can receive $m_k$ measurements $z^1_k, z^2_k,...,z^{m_k}_k$ at each time step $k$ each taking values in the measurement space $\mathcal{Z}$, where $m_k$ and $z^i_k, \forall ~i$ are random and time varying. Thus the group of received measurements by an agent at time $k$ can be modeled as the set:
\begin{equation}
     Z_k = \{z^1_k, z^2_k,...,z^{m_k}_k\} ~ \in \mathcal{F}(\mathcal{Z}) 
\end{equation}
where $\mathcal{F}(\mathcal{Z})$ denotes the space of all finite subsets of $\mathcal{Z}$. In Sec. \ref{sec:Background}-\ref{sec:Problem_Formulation} it will be shown that $Z_k$ is also a random finite set.  $Z_k$ will be referred to as the \textit{multi-target measurement set}. 

\begin{figure}
	\centering
	\includegraphics[width=\columnwidth]{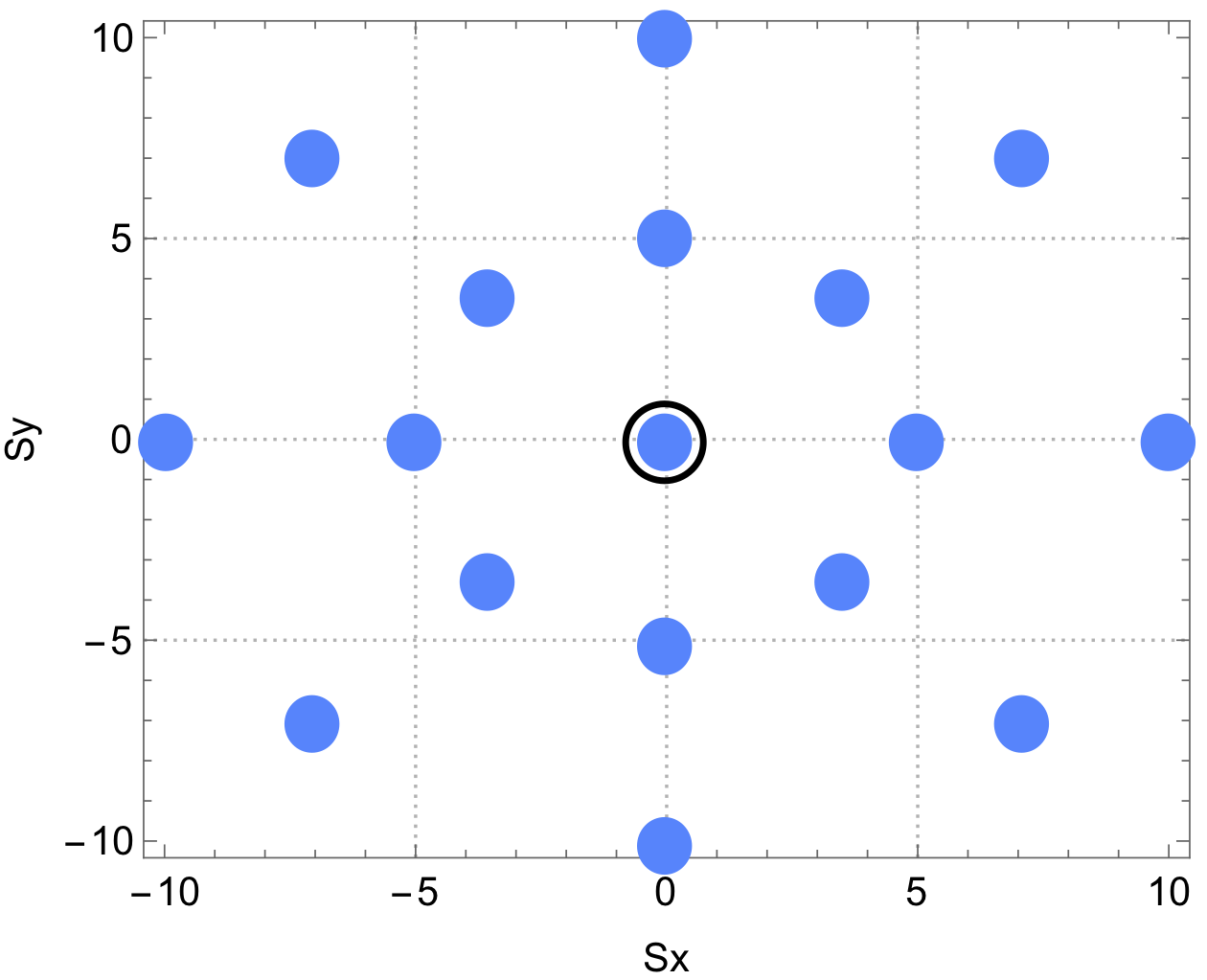}
	\caption{The figure shows an illustrative example of the agent's admissible control actions $\mathbb{U}_k$ at time $k$. In this example the radial displacement $\Delta_R=5$m, $N_R=2$ and $N_\theta = 8$ which gives a total of 17 control actions, including the initial position of the agent at time $k-1$ which is marked with a circle.}
	\label{fig:controls}
\end{figure}

\subsection{Agent Dynamics}
\label{ssec:AgentDynamics}
Let $S = \{1,2,...,|S|\}$ be the set of all mobile agents that we have in our disposal operating in a discrete-time setting.
At time $k$ the 2D surveillance area $\mathcal{A} \subseteq \mathbb{R}^2$ is monitored by $|S|$ mobile agents with states $s^1_k,s^2_k,...,s^{|S|}_k$, each taking values in $\mathcal{A}$. Each agent is subject to the following dynamics with bounded control inputs:

\begin{equation} \label{eq:controlVectors}
s^j_{k} = s^j_{k-1} + \begin{bmatrix}
						l_1\Delta_R \text{cos}(l_2 \Delta_\theta)\\
						l_1\Delta_R \text{sin}(l_2 \Delta_\theta)
					\end{bmatrix},  
					\begin{array}{l} 
						l_2 = 0,...,N_\theta\\ 
						l_1 = 0,...,N_R
				    \end{array} 
\end{equation}
where  $s^j_{k-1} = [s^j_x,s^j_y]^\top_{k-1}$ denotes the position (i.e. xy-coordinates) of the $j_{\text{th}}$ agent at time $k-1$, $\Delta_R$ is the radial step size, $\Delta_\theta=2\pi/N_\theta$ and the parameters $(N_\theta,N_R)$ control the number of possible control actions. We denote the set of all admissible control actions of agent $j$ at time $k$ as $\mathbb{U}^j_{k}=\{s^{j,1}_{k},s^{j,2}_{k},...,s^{j,|\mathbb{U}_{k}|}_{k}  \}$ as computed by Eqn. (\ref{eq:controlVectors}). An illustrative example is shown in Fig. \ref{fig:controls}.

\subsection{JoSAT Objective}
The jointly optimized search and track objective can now be more formally defined as:
\begin{equation} \label{eq:o1}
(\text{P}_1)  \underset{u^j_k, \forall j \in S}{\arg\min} \left[ \xi^\alpha_\text{search}(\{u^j_k : j \in \alpha\}) + \underset{\substack{j=1 \\ j \not\in \alpha }}{\sum^{|S|}} \xi^j_\text{track}(u^j_{k},Z^j_{k|k-1,u^j_{k}})  \right]
\end{equation}

\begin{equation*}
\begin{aligned}
& \text{subject to} & & \alpha \in \mathbb{P}(S) \\
&&& u^j_{k} \in \mathbb{U}^j_{k} ~,\>\forall j \in S\\
&&& Z^j_{k|k-1,u^j_{k}} = Z_\text{PIMS}(\hat{X}^j_{k|k-1},u^j_{k})~,\>\forall j \in S\\
&&& \norm{u_k^i - u_k^j}> d_\text{min}~, \forall ~i \ne j
\end{aligned}
\end{equation*}

\noindent where $\alpha$ is the decision variable for the agents' operating mode and $\mathbb{P}(S) =\Big \{ \emptyset,\{1\},\{2\},\dots,\{|S|\},\{1,2\},\dots \Big \}$ denotes the power set of $S$. The function $\xi^\alpha_\text{search}(\{u^j_k : j \in \alpha\})$ takes as input a set of control actions for the agents in search mode and returns the total cost of \text{searching} and the function $\xi^j_\text{track}(u^j_{k},Z^j_{k|k-1,u^j_{k}})$ returns the cost of tracking when the control $u^j_k$ is taken and the predicted multi-target measurement set $Z^j_{k|k-1,u^j_{k}}$ is supposed to have been received by agent $j$. $\hat{X}^j_{k|k-1}$ denotes the estimated multi-target state for time $k$ from agent $j$ based on time $k-1$. 

The cost of tracking depends on the control $u^j_k$ that is applied to agent $j$ through the future received measurements (i.e. depending on the control action taken, different measurements will be received). 
The measurement generator function $Z_\text{PIMS}(\hat{X}^j_{k|k-1},u^j_{k})$ generates the predicted ideal multi-target measurement set based on $\hat{X}^j_{k|k-1}$ and $u^j_k$.

 Finally, the constraint $\norm{u_k^i - u_k^j}> d_\text{min}, \forall ~i \ne j$ does not allow any two agents to get close to each other. This is by design since in this work there is no benefit of two agents to be close to each other, and for instance search the same area or perform tracking of the same target. 
 
 We should note here that the optimization of Eqn. (\ref{eq:o1}) finds the joint control actions over all agents, and at the same time partitions the agents into two groups i.e. search or track so that the joint search-and-track cost is minimum. In other words, at each time-step, the system automatically assigns the agents to searching or tracking mode and selects the joint control actions which minimize the JoSAT objective. Also note, that in this work we do not assume any time delays between agents i.e. all agents have access to the same clock and they are fully synchronized.

\section{System overview} \label{sec:Sys_Arch}
\begin{figure*}
	\centering
	\includegraphics[width=\textwidth]{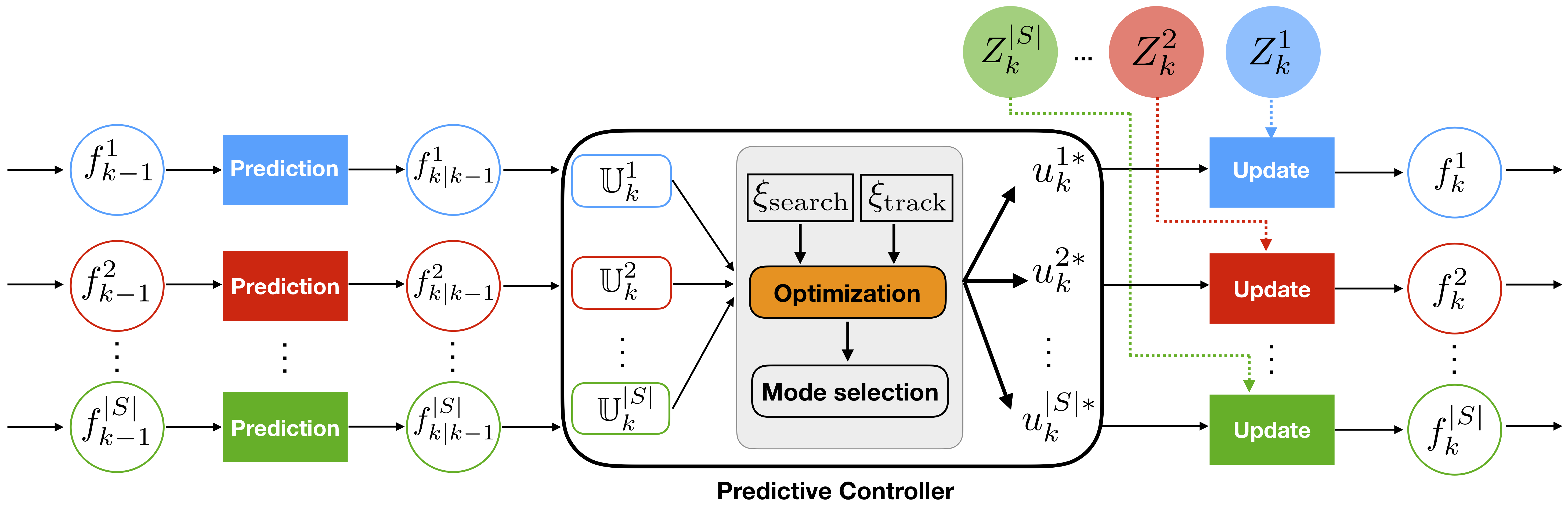}
	\caption{The figure illustrates the proposed system architecture with $S = \{1,2,...,|S|\}$ agents. The following recursion is shown: a) for each agent $j$ the predictive multi-object probability distribution $f^j_{k|k-1}$ is first computed from the posterior $f^j_{k-1}$ using the prediction step, b) the estimated numbed of targets and their states are extracted from each predictive multi-object distribution, c) an optimization problem is then solved where different hypothetical control actions from the set of admissible control actions $\mathbb{U}^j_{k}$ are generated for each agent. The mode of operation i.e. \textit{search} or \textit{track} and the corresponding control actions for all agents are  jointly selected so that the total cost of searching (given by the searching objective function i.e. $\xi_\text{search}$) and tracking (given by the tracking objective function i.e. $\xi_\text{track}$) is minimized. d) Finally, the agents move to the new states according to the selected optimized controls $\{ u^{1*}_k,\dots,u^{|S|*}_k \}$ and compute the posterior multi-object distribution $f^j_k$ using the received measurements $Z^j_k$ at time $k$. The same procedure is the repeated for the next time instance.}	
	\label{fig:sys_arch}
\end{figure*}



In this section we give an overview of the proposed system architecture, shown in Fig. \ref{fig:sys_arch}, and we outline how we have incorporated the theory of multi-object filtering to the problem of multi-agent joint searching and tracking.

To summarize, an agent $j \in S$ can operate in either \textit{search} or \textit{track} mode and in each mode the agent is subject to a set of admissible control actions denoted by $\mathbb{U}^j_{k}$ for time $k$. The agents can dynamically switch between the two modes during the mission in order to maximize the joint global objective of searching and tracking i.e. Eqn (\ref{eq:o1}). When a subset of agents $\alpha \subseteq S$ is in \textit{search} mode the collective objective over all agents $j \in \alpha$ is the maximization of the covered area. In other words the agents should traverse the surveillance area in such a way so that the coverage is maximized. On the other hand when a subset of agents $\alpha^\text{C} \subseteq S$ (where $\alpha^\text{C}$ denotes the complement of $\alpha$ with respect to $S$) is in \textit{track} mode, their objective is the maximization of the tracking performance, i.e. to move in locations where the number of targets and their states is estimated as accurately as possible. Deciding on the subset of agents in \textit{search} or \textit{track} mode is governed by the respective objective functions that control the movement of the agents to achieve the best tradeoff. In other words the proposed controller solves the optimization of Eqn. (\ref{eq:o1}).
 
In order to achieve this, the proposed system takes the following steps: The number of targets and their states at time $k$ are jointly modeled as a random finite set $X_k$ (also known as \textit{multi-target state}) which is a generalization of a random variable/vector (see Sec. \ref{sec:Background}). 

Each agent maintains a \textit{multi-object} probability distribution (see Sec. \ref{sec:Background}-\ref{sec:Problem_Formulation}) which jointly accounts for the uncertainty in the number of targets and their states (i.e. position and velocity) inside its sensing range (this multi-object probability distribution is a generalization of the probability density function on random finite sets i.e. $X$). Let the posterior \textit{multi-object} distribution at time $k-1$ be denoted by $f^j_{k-1}$ for agent $j$. To be precise $f^j_{k-1} = f^j_{k-1}(X^j_{k-1}|Z^j_{1:k-1})$ where $X^j_{k-1}$ is conditioned on the received measurement sets $Z^j_{1:k-1}$ up to time $k-1$, however we will often use the short notation i.e. $f^j_{k-1}$ to denote this distribution.

In the first step the predictive density $f^j_{k|k-1}=f^j_{k|k-1}(X^j_{k}|Z^j_{1:k-1})$ is obtained through a prediction step as the figure shows. This allows us to make predictions regarding the number of targets and their states for the next time-step. The estimated number of targets ($\hat{n}^j_{k|k-1}$) and their states ($\hat{X}^j_{k|k-1}$) for the next time-step are extracted from the predictive density.

 Based on these predictions the proposed controller finds the combination of control actions $(u^j_k \in \mathbb{U}_k^j ~ \forall j)$ as well as the mode of each agent which minimizes the convex sum of the search and track objective functions (see Sec. \ref{sec:Problem_Formulation}).
 
The optimal control actions $u^{1*}_k,...,u^{|S|*}_k$ are then applied to the agents. Each agent then moves to its new state at time $k$ where it receives the multi-target measurement set $Z^j_k$ (if it exists). This measurement set at time $k$ is then used in the update step i.e. each agent uses the received measurement set to compute the posterior distribution $f^j_k = f^j_{k}(X^j_{k}|Z^j_{1:k})$. The updated (i.e. final) estimates of the number of targets $\hat{n}^j_{k}$ and their states $\hat{X}^j_{k}$ for time $k$ are then computed from $f^j_k$ for each agent $j$. The total estimated number of targets at time $k$ is then computed as $\hat{n}_k = \sum_j \hat{n}^j_k, \forall j$ and their states as $\hat{X}_k = \bigcup_j \hat{X}^j_k, \forall j$. This procedure is repeated recursively over time. 

We should note here that the number of targets (and their states) is being estimated recursively over time. Based on these estimates, the mode of operation should be carefully chosen. For instance when no targets are inside the surveillance area, all agents should be in \textit{search} mode, etc. It is also worth noting that the control actions taken, affect the received measurements which in turn affect the estimation of the target state during the update step. Hence, to optimize the control actions would require the knowledge of all the future measurements. In the next section we provide an overview of random finite sets for multiple target tracking which we use subsequently to address the aforementioned challenge in our proposed approach. 

\section{Background on Multi-Target Tracking} \label{sec:Background}
Multiple target tracking (MTT) \cite{Blackman1999,Mahler2014book} refers to the problem of estimating the number and states of multiple targets from noisy sensor measurements in the presence of false alarms or clutter. MTT is nowadays found in many applications including surveillance, defense, autonomous vehicles, robotics, etc. In this section we only give a brief overview on the various MTT techniques. A more detailed description of MTT algorithms can be found in \cite{Blackman1999,VoBook2015}. 

In this work we divide MTT algorithms in two main categories namely data association-based and data association free. Data association-based MTT algorithms such as the MHT, JPDA and GNN \cite{Blackman1999,VoBook2015} require to first solve the measurements-to-tracks assignment problem and only then proceed with the multi-target state estimation. On the other hand, data association free methods such as the RFS-based PHD, CPHD and Multi-Bernoulli filters \cite{Mahler2007book,Mahler2014book} can bypass the data association problem and proceed directly with the multi-target state estimation. Depending on the application, this property could be highly desirable since it significantly reduces the computational complexity of MTT algorithms. 

The PHD filter \cite{Mahler2003} is the first computationally desirable approximation of the multi-target Bayes posterior. In particular it propagates the first-order statistical moment of the multi-target posterior in place of the full posterior distribution. The computationally more expensive CPHD filter \cite{Mahler2007} provides a better performance compared to the PHD filter since it jointly propagates the first-statistical moment and the cardinality distribution. Unlike the PHD and CPHD filters which propagate moments the multi-Bernoulli filter \cite{Vo2009} approximates the true multi-target posterior distribution as multi-Bernoulli and propagates the parameters of a multi-Bernoulli distribution. The multi-Bernoulli filter avoids many of the problems present in the PHD/CPHD filters i.e. these filters require a clustering step to extract state estimates which introduces an additional source of error and increases the computationally complexity. This makes the multi-Bernoulli filter a more attractive solution with linear complexity in the number of targets and measurements.

 We should also point out that with the introduction of labeled RFSs \cite{Vo2011lrfs,Vo2013lrfs} and in particular with the inception of the generalized labeled multi-Bernoulli (GLMB) density the multi-target Bayes filter can tackle the data-association problem and provide target tracks. More specifically, the GLMB multi-target filter \cite{Vo2014lrfs} is able to simultaneously estimate the number of targets and their states from a set of noisy observations in the presence of data association uncertainty, detection uncertainty and clutter. However, compared to the multi-Bernoulli filter the computational complexity of the GLMB filter is at best cubic in the number of measurements. An extension to the multi-Bernoulli filter i.e. the Labeled multi-Bernoulli (LMB) filter, was presented in \cite{Reuter2014lrfs}. The LMB filter achieves better performance compared to the multi-Bernoulli filter while at the same time outputs target tracks. This however comes with an increased computational cost i.e. its computational complexity is at worst cubic in the number of measurements. Recently the authors in \cite{Vo2017lrfs} have presented a more efficient implementation of the GLMB filter with linear complexity in the number of measurements and quadratic in the number of targets by combining the prediction and update into a single step.

\subsection{Random Finite Sets}
A random finite set (RFS) is a finite-set-valued random variable, which differs from a random vector in two ways: a) the number of elements in a RFS is random and b) the order of the elements in a RFS is irrelevant. More specifically, a RFS $X$ is completely specified by a) its cardinality distribution $\rho(n) = p(|X|=n),~ n \in \mathbb{N}_0$ which defines a probability distribution over the number of elements in $X$ and b) by a family of conditional joint symmetric probability distributions $p(x_1,...,x_n|n)$,~ $x_1,...,x_n \in \mathcal{X}$ that characterize the distribution of its elements over the state space $\mathcal{X}$. In connection with our problem, the random variable $n$ can be used to model the time-varying number of targets (i.e. the number of survivors at some point in time) and thus $\rho(n)$ is the probability distribution of the number of targets.

The belief density or otherwise known as the \textit{multi-object} probability density function (pdf) $f(X)$ of the RFS $X$ is given by: $f(X) = f(\{x_1,...,x_n\}) = n!  \rho(n) p(x_1,...,x_n|n)$ and the notion of integration is given by the set-integral which is defined as:
\begin{equation}
	\int_{\mathcal{F}(\mathcal{X})} f(X) \delta X = f(\emptyset)+\sum^{\infty}_{n=1} \frac{1}{n!} \int f(\{x_1,...,x_n\})d_{x_1}...d_{x_n}
\end{equation}
\noindent where $\mathcal{F}(\mathcal{X})$ is the space of finite subsets of $\mathcal{X}$. The following RFSs are relevant in this paper:
\subsubsection{Bernoulli RFS} The Bernoulli RFS $X$ can either be empty with probability $1-r, ~ r \in (0,1)$ or be a singleton set (i.e. its set cardinality is equal to one) with (existence) probability $r$ and with its element distributed over the state space $\mathcal{X}$ according to pdf $p(x)$. The Bernoulli multi-object pdf is given by:
\begin{equation}
 f(X) = 
  \begin{cases} 
   1-r & \text{if } X = \emptyset \\
   r p(x) & \text{if } X=\{x\}
  \end{cases}
\end{equation}
Thus a Bernoulli RFS can be completely characterized by the parameter set $(r,p(x))$.
\subsubsection{Multi-Bernoulli RFS} The multi-Bernoulli RFS $X = \bigcup^v_{i=1} X_i$ is a union of a fixed number $v$ (where $v$ is known) of independent Bernoulli RFSs $X_i$ with parameters $\{(r^i, p^i(.))\}^{v}_{i=1}$ with multi-object pdf given by (see \cite{Mahler2014book} pp. 102):
\begin{equation}\label{eq:multi-bernoulli}
	f(\{x_1,...,x_n\}) = \underset{1 \le i_1 \ne,...,\ne i_n \le v}{\sum} \Big( Q_{i_1,...,i_n} \cdot p^{i_1}(x_1) \cdot \cdot \cdot p^{i_n}(x_n) \Big)
\end{equation}
where 
\begin{equation*}
Q_{i_1,...,i_n} = \left( \prod_{i=1}^v (1-r^i) \right) \cdot \frac{r^{i_1}}{1 -r^{i_1}}\cdot\cdot\cdot \frac{r^{i_n}}{1 -r^{i_n}}
\end{equation*}

\noindent Here the notation $\sum_{1 \le i_1 \ne,...,\ne i_n \le v}(.)$ enumerates the $n$-permutations of $v$ i.e. $P^v_n =\frac{v!}{(v-n)!} $. In other words, it enumerates all ordered arrangements of $n$ elements taken from a set of size $v$. A multi-Bernoulli distribution is completely specified with the parameter set $\{(r^i,p^i(.)) \}_{i=1}^v$.
In connection with our problem $v$ can be used to model the maximum number of hypothetical targets maintained by an agent.

\subsubsection{Poisson RFS} The poisson RFS $X$ has a cardinality distribution which is Poisson with parameter $\lambda$ i.e. $\rho(n) = \frac{e^{-\lambda}\lambda^n}{n!}, ~n =0,1,2... $ and elements which are independent and identically distributed (IID) random variables and distributed according to $p(x)$ on $\mathcal{X}$. The multi-object pdf is given by: 
\begin{equation}
    f(X) = e^{-\lambda}\prod_{x \in X} \kappa(x)
\end{equation}
\noindent where $\kappa(x) = \lambda ~ p(x)$ is called the intensity function for the Poison RFS.

\subsection{Multi-object Stochastic Filtering}
In stochastic filtering \cite{Simon2006} we are interested in the posterior probability density $p_k(x_k|z_{1:k})$ of some hidden state $x_k$ at time $k$ given all measurements $z_{1:k}=z_1,...,z_k$. Assuming an initial density on the state $p_0(x_0)$, the posterior density at time $k$ can be computed using the Bayes recursion as:
\begin{align} 
p_{k|k-1}&(x_k|z_{1:k-1})= \label{eq:predict}\\
& \int  \pi_{k|k-1}(x_k|x_{k-1}) ~ p_{k-1}(x_{k-1}|z_{1:k-1}) ~d x_{k-1} \notag
\end{align}
\vspace{-6mm}
\begin{align}
p_k(x_k|z_{1:k}) = \frac{g_k(z_k|x_k) ~ p_{k|k-1}(x_k|z_{1:k-1})}{\int g_k(z_k|x_k) ~ p_{k|k-1}(x_k|z_{1:k-1}) ~dx_k} \label{eq:update}
\end{align}
where Eqn. (\ref{eq:predict}) and (\ref{eq:update}) are referred to as the prediction and update steps respectively. The function $\pi_{k|k-1}(x_k|x_{k-1})$ models the uncertainty of the current state given the previous state and is referred to as the transitional density and the function $g_k(z_k|x_k)$ describes the distribution of measurements given the current state and is referred to as the measurement likelihood function. 
At each time step the hidden state is usually extracted from the posterior distribution using the expected a posteriori (EAP) or the maximum a posteriori (MAP) estimators.

The above recursion can be extended \cite{Mahler2007book} in the case of RFSs.  Suppose now that the hidden state and measurements are RFS and that at time $k-1$ the posterior multi-object pdf of $X$ is given by $f_{k-1}(X_{k-1}|Z_{1:k-1})$. The predicted and updated multi-object densities are then given by the multi-object Bayes recursion:

\begin{align} 
f_{k|k-1}&(X_k|Z_{1:k-1}) \label{eq:predictRFS}\\
&= \int  \phi_{k|k-1}(X_k|X_{k-1}) ~ f_{k-1}(X_{k-1}|Z_{1:k-1}) ~\delta X_{k-1} \notag
\end{align} 
\vspace{-2mm}
\begin{align}
f_k(X_k|Z_{1:k}) &= \frac{\varphi_k(Z_k|X_k) ~ f_{k|k-1}(X_k|Z_{1:k-1})}{\int \varphi_k(Z_k|X_k) ~ f_{k|k-1}(X_k|Z_{1:k-1}) ~\delta X_k} \label{eq:updateRFS}
\end{align}

\noindent where the integrals in Eqn. (\ref{eq:predictRFS})-(\ref{eq:updateRFS}) are set integrals, and the functions $\phi_{k|k-1}(X_k|X_{k-1})$ and $\varphi_k(Z_k|X_k)$ are the multi-object transitional density and the multi-object likelihood function respectively. Because the recursion above has no analytic solution in general, several approximations exist (as in \cite{Mahler2014book}) which propagate only summary statistics instead of the full posterior density. A prominent approach is the multi-Bernoulli filter \cite{Vo2009,Vo2007multi} which is used in this work and elaborated in Sec. \ref{sec:Problem_Formulation}.

\section{JoSAT Multi-Target Framework}
\label{sec:Problem_Formulation}
This section develops the components of the JoSAT system using random finite sets and derives the multi-agent objective functions for the tasks of searching and tracking. Intuitively, the collection of targets at each time instance can be seen as a set which changes size as targets appear and disappear from the surveillance region (e.g. due to births and deaths). Since this set consists of a random number of random variables it can be modeled as a random finite set and the same is true for the received measurements as we detail below.

\subsection{Multi-target Dynamics}\label{ssec:TargetDynamics}
Earlier in subsection \ref{ssec:single_target_dynamics} we have described the single target dynamics, in this subsection we will use random finite sets to describe the multi-target dynamics. 

Using the RFS theory we can now define a multi-object dynamic model i.e. the evolution in time of the RFS $X_k$ as follows:
\begin{equation}\label{eq:RFS_transition_model}
    X_k = \left[ \underset{x_{k-1} \in X_{k-1}}{\bigcup} \Psi(x_{k-1}) \right] \cup \Gamma_k
\end{equation}
where $\Psi(x_{k-1})$ is a Bernoulli RFS which models the evolution of the set from the previous state, with parameters $(p_{S}(x_{k-1}),\pi_{k|k-1}(x_k|x_{k-1}))$. Thus a target with state $x_{k-1}$ continues to exists at time $k$ with surviving probability $p_{S}(x_{k-1})$ and moves to a new state $x_k$ with transition probability $\pi_{k|k-1}(x_k|x_{k-1})$, (see Eqn. (\ref{eq:single_dynamics})). Otherwise the target dies with probability $1-p_{S}(x_{k-1})$. Note here that $X_k$ is a multi-Bernoulli RFS assuming that the RFSs constituting the union in Eqn. (\ref{eq:RFS_transition_model}) are mutually independent. Finally, the term $\Gamma_k$ denotes the multi-Bernoulli RFS of spontaneous births with uniform spatial distribution over $\mathcal{X}$. We now have a multi-object transition model i.e. Eqn. (\ref{eq:RFS_transition_model}) which jointly incorporates motion, birth and death for multiple targets.     

\subsection{Multi-target Measurement Model}\label{ssec:AgentSensing}                                                                               
At each time step an agent can receive a time-varying number of measurements i.e. the number of measurements is random and depends on: a) the number of targets inside the agent's sensing range which is random as well, b) whether all targets have been detected or not, and c) whether measurements come from targets or from clutter (i.e. false alarms). That said, the measurements received at some time instance can be modeled as a random finite set. The multi-target measurement set $Z^j_k$ of agent $j$ at time $k$ is given by:
\begin{equation}\label{eq:RFS_measurement_model}
    Z^j_k = \left[ \underset{x^j_{k} \in X^j_{k}}{\bigcup} \Theta(x^j_{k}) \right] \cup \text{K}_k
\end{equation}
where $X^j_k \subseteq X_k$ is the multi-target state perceived by the $j_\text{th}$ agent. $\Theta(x^j_{k})$ is a Bernoulli RFS which models the target generated measurements with parameters $(p_{D}(x^j_k,s^j_k),g_k(z^{ji}_k|x^j_k,s^j_k))$. Thus a target with state $x^j_k$ at time $k$ is detected by the $j_\text{th}$ agent with state $s^j_k$ with probability $p_{D}(x^j_k,s^j_k)$ (see Eqn. (\ref{eq:sensing_model})) and generates a measurement $z^{ji}_k$ with likelihood $g_k(z^{ji}_k|x^j_k,s^j_k)$ (see Eqn. (\ref{eq:meas_model})) or is missed with probability $1-p_{D}(x^j_k,s^j_k)$ and generates no measurements. Additionally an agent can receive false alarms measurements i.e. the term $\text{K}_k$ is a Poisson RFS which models the set of false alarms or clutter received by an agent at time $k$ with intensity function $\kappa_k(z^{ji}_k) = \lambda f_{c}(z^{ji}_k)$, where in this paper $f_c(.)$ denotes the uniform distribution over $\mathcal{Z}$. 

Note here that Eqn. (\ref{eq:RFS_measurement_model}) accounts for the detection uncertainty and clutter and that in this work the following holds by design $X_k = \bigcup_j X^j_k, \forall j$ and $X^{j_1}_k \cap X^{j_2}_k = \emptyset, ~ \forall j_1, j_2$ i.e. a target cannot be tracked by more than one agent. 

\subsection{Tracking Multiple Targets}
When the agent  $j$ is in \textit{track} mode the objective is to estimate at each time $k$ both the number of targets $n^j_k$ and their respective states $x_k^{ji}, \forall i \in \{1,...,n^j_k\}$ inside its sensing range from a sequence of noisy measurement sets $Z^j_{1:k}$. In essence we would like to compute the multi-object density of $X^j_k$ given the measurement sets $Z^j_{1:k}$ i.e. $f^j_{k}(X^j_{k}|Z^j_{1:k})$ using the multi-object Bayes recursion of Eqn. (\ref{eq:predictRFS}) - (\ref{eq:updateRFS}). The RFS formulation of subsections \ref{ssec:TargetDynamics}-\ref{ssec:AgentSensing} however, allows us to compute this recursion with a multi-Bernoulli filter \cite{Vo2009} which approximates the multi-object posterior density as multi-Bernoulli distribution and propagates only its parameters $\{(r^{ji}_k, p^{ji}_k(.))\}^{v_k^j}_{i=1}$ in time instead of the full distribution. For brevity, in the rest of the paper, we will denote the multi-Bernoulli distribution at time $k$ as $f_{k}= \{(r^i_{k},p^i_{k}(.))\}^{v_{k}}_{i=1}$.

Furthermore, the multi-Bernoulli filter recursion is summarized here. A detailed description of the multi-Bernoulli filter can be found in \cite{Vo2009}. 
The multi-Bernoulli filter proceeds as follows: The predictive multi-object density at time $k$ is given by:
\begin{equation} \label{eq:RFS_prediction1}
	f_{k|k-1} = f^\text{persist}_{k|k-1} \cup f^\text{birth}_{k}
\end{equation}
where $f^\text{persist}_{k|k-1} = \{(r^i_{k|k-1},p^i_{k|k-1}(.))\}^{v_{k-1}}_{i=1}$ describes the set of persisting targets from the previous time-step and $f^\text{birth}_{k} = \{(r^i_{k},p^i_{k}(.))\}^{\Gamma_{k}}_{i=1}$ are the parameters of a multi-Bernoulli RFS for the new-born targets at time $k$. $f^\text{persist}_{k|k-1}$ is further given by:
\begin{align}
	r^i_{k|k-1} &= r^i_{k-1}\left< p^i_{k-1} , p_{S}\right>  \label{eq:RFS_prediction2}\\
	p^i_{k|k-1}(x) &= \frac{\left< \pi_{k|k-1}(x|.) , p^i_{k-1}(.)  p_{S}(.) \right>}{\left< p^i_{k-1} , p_{S}\right>} \label{eq:RFS_prediction3}
\end{align}
where the notation $<f,g>$ is defined as $\int f(x)g(x)dx$. 

Let the predictive multi-object density denoted by $f_{k|k-1} = \{(r^i_{k|k-1},p^i_{k|k-1}(.))\}^{v_{k|k-1}}_{i=1}$ where now the term $v_{k|k-1}$ denotes the number of hypothesized Bernoulli components after accounting for the target births and let the multi-object measurement set at time $k$ be $Z_k=\{z^1_k,...,z^{m_k}_k\}$. The posterior multi-object distribution at time $k$ is given by a multi-Bernoulli distribution with parameters:
\begin{equation} \label{eq:RFS_update1}
	f_k = f^\text{legacy}_{k} \cup f^\text{meas}_{k}
\end{equation}
where $f^\text{legacy}_{k} = \{(r^i_{k},p^i_{k}(.))\}^{v_{k|k-1}}_{i=1}$ contains legacy tracks i.e. updates of predicted tracks, assuming that no measurements were collected from them and given by:
\begin{align}
	r^i_{k} &= r^i_{k|k-1} \frac{1 - \left< p^i_{k|k-1} , p_{D}(.,s_k)\right>}{1 - r^i_{k|k-1}\left< p^i_{k|k-1} , p_{D}(.,s_k)\right>} \label{eq:r_update_legacy} \\
	p^i_{k}(x_k) &= p^i_{k|k-1}(x_k) \frac{1-p_{D}(x_k,s_k)}{1 - \left< p^i_{k|k-1} , p_{D}(.,s_k)\right>} \label{eq:p_update_legacy}
\end{align}

\noindent where $p_{D}(x_k,s_k)$ is the probability of detecting a target as described in subsection \ref{ssec:measM}.
$f^\text{meas}_{k} = \{(r_{k}(z),p_{k}(.,z))\}_{z \in Z_k}$ contains new tracks i.e. joint updates of all predicted tracks using each of the measurements separately; and given by:
\begin{align}
	r_{k}(z) &= \frac{\sum \displaylimits^{v_{k|k-1}}_{i=1} \frac{r^i_{k|k-1}(1-r^i_{k|k-1})\left< p^i_{k|k-1} , L^z_k\right> }{\left (1-r^i_{k|k-1} \left<p^i_{k|k-1} , p_{D}(.,s_k)\right> \right)^2}}{\kappa_k(z) + \sum \displaylimits^{v_{k|k-1}}_{i=1} \frac{r^i_{k|k-1}\left< p^i_{k|k-1} , L^z_k\right>}{1-r^i_{k|k-1} \left<p^i_{k|k-1} , p_{D}(.,s_k)\right>}} \label{eq:r_update_meas}\\
	p_{k}(x_k,z) &= \frac{\sum \displaylimits^{v_{k|k-1}}_{i=1} \frac{r^i_{k|k-1}}{1-r^i_{k|k-1}} p^i_{k|k-1} L^z_k(x_k)}{\sum \displaylimits^{v_{k|k-1}}_{i=1} \frac{r^i_{k|k-1}}{1-r^i_{k|k-1}} \left< p^i_{k|k-1} \cdot L^z_k\right> } \label{eq:p_update_meas}
\end{align}
where $L^z_k(x_k) = g_k(z_k|x_k) p_{D}(x_k,s_k)$.

The posterior multi-object density at time $k$ for agent $j$ is thus given by $f^j_k = \{(r^{ji}_{k},p^{ji}_{k}(.))\}^{v^j_k}_{i=1}$ from where the estimated number of targets $\hat{n}^j_k$ and the multi-target state $\hat{X}^j_k$ can be computed (see Sec. \ref{ssec:trackObjective}). The total estimated number of targets at time $k$ in the surveillance area is then computed as $\hat{n}_k = \sum_j \hat{n}^j_k, \forall j$ and the multi-target state which accounts for all targets in the area as $\hat{X}_k = \bigcup_j \hat{X}^j_k, \forall j$.

\subsection{Multi-Agent Tracking Objective} \label{ssec:trackObjective}
Let the multi-Bernoulli distribution $f^j_{k} = \{(r^{ji}_{k}, p^{ji}_{k}(.))\}^{v^j_{k}}_{i=1}$ from agent $j$ where $v^j_k$ is 
the maximum number of hypothesized targets, $r^{ji}_k$ denotes the probability of existence of the $i_\text{th}$ hypothesized target 
(i.e. how likely the $i_\text{th}$ hypothesized target is a true target) and $p^{ji}_k(.)$ denotes the corresponding posterior 
density of the state of the $i_\text{th}$ target. 

The expected a posteriori (EAP) estimate of number of targets $\hat{n}^j_k$ can be computed as 
$\hat{n}^j_k = round(\sum_i r^{ji}_k)$ and the multi-target state $\hat{X}^j_k$ can then be computed by 
taking the $\hat{n}^j_k$ components with the highest probabilities of existence and extracting the single target 
states (i.e. $\hat{x}_k$) from their individual posterior 
densities as $\hat{X}^j_k = \bigcup_{i=1}^{\hat{n}^j_k} \{\hat{x}_k~=~\arg\max~p^{ji}_k(x_k)\}$.

In tracking, we wish to estimate as accurately as possible the $\hat{n}^j_k$ and $\hat{X}^j_k$ for all $j$ agents. 
Intuitively, one way to achieve this is to minimize the variance in the posterior distribution $f^j_k$ 
from which we derive the above estimates. In this work we build upon this and we denote the tracking objective 
function we wish to minimize by $\xi^j_\text{track}(u^j_{k},Z^j_{k}),~ u^j_{k} \in \mathbb{U}^j_{k}$ if we were 
going to apply at time $k$ the control action $u^j_{k}$ and subsequently observe at the same time-step the 
measurement set $Z^j_{k}$. Note here that the received measurement set $Z^j_k$ depends on $u^j_k$ 
(i.e. the control action taken) by the agent $j$ through Eqn. (\ref{eq:sensing_model}) - (\ref{eq:meas_model}). 
Also observe that the objective function we wish to minimize depends on an unknown future measurement set, 
i.e. the measurement set $Z^j_k$ will be received once the control $u^j_k$ has been determined and applied. 
In other words, this objective function can be evaluated after the control action $u^j_{k}$ has been taken, 
because only then the measurement $Z^j_k$ becomes available.  
  
A common approach \cite{Clark2011,Beard2015} around this problem is to take the statistical 
expectation of the objective function with respect to all possible values of 
measurements i.e. $\mathbb{E} [\xi^j_\text{track}(u^j_{k},Z^j_{k})]$ thus the optimization problem 
becomes $u_{k}^{j\star} = \arg\min~ \mathbb{E} \left[ \xi^j_\text{track}(u^j_{k},Z^j_{k}) \right]$. 
However, computing this expectation is computationally intensive because it requires the generation of an ensemble  
of measurement sets (pseudo-measurements) $Z^j_{k}$ for each hypothesized control action. 

An alternative and computationally cheaper approach which we adopt in the paper uses the predicted ideal measurement set (PIMS)\cite{Mahler2004} i.e. the noise-free clutter-free measurement set which is most likely to be obtained when a particular control action is applied. We denote the PIMS measurement set as $Z^j_{k|k-1}$ and thus the control problem now becomes:

\begin{equation} \label{eq:optimize_uk}
	u_{k}^{j\star} = \underset{u^j_{k} \in \mathbb{U}^j_{k}}{\arg\min}~ \xi^j_\text{track}(u^j_{k},Z^j_{k|k-1})
\end{equation}

\noindent To optimize Eqn. (\ref{eq:optimize_uk}) let the multi-Bernoulli distribution at time $k-1$ be given by $f^j_{k-1} = \{(r^{ji}_{k-1}(u^j_{k-1}), p^{ji}_{k-1}(.,u^j_{k-1}))\}^{v^j_{k-1}}_{i=1}$ where now by using this notation we explicitly show its dependence on the control action that was taken for the time-step $k-1$. At time $k$ the agent has available a set of admissible control actions $u^j_{k} \in \mathbb{U}^j_{k}$ given by Eqn. (\ref{eq:controlVectors}), one of which should be selected in order to take the agent to the next state at time $k$. First we compute the multi-Bernoulli predictive density  $f^j_{k|k-1}=\{(r^{ji}_{k|k-1}(u^j_{k-1}),p^{ji}_{k|k-1}(.,u^j_{k-1}))\}^{v^j_{k|k-1}}_{i=1}$ which is obtained without yet performing any control action. Then we apply a pre-estimation step where the number of targets $\hat{n}^j_{k|k-1}$ is first estimated from the predictive density by counting the Bernoulli components which exhibit a probability of existence $r^{ji}_{k|k-1}>0.5, \forall i \in \{1...v^j_{k|k-1}\}$. The states of those targets are then extracted from their spatial distributions to obtain $\hat{X}^j_{k|k-1}$. Then for a given hypothesized control action $u^j_{k} \in \mathbb{U}^j_{k}$ the corresponding PIMS $Z^j_{k|k-1  ,u^j_k}$ is generated as:
\begin{align} \label{eq:GenPIMS}
     Z^j_{k|k-1,u^j_k} &= Z^j_{k|k-1,u^j_k} ~~\cup \\
      &\{ \underset{z}{\arg\max}~ g_k(z|\hat{x}^j_{k|k-1},u^j_k)\}, ~ \forall \hat{x}^j_{k|k-1} \in \hat{X}^j_{k|k-1} \notag
\end{align}

Suppose now that for each pair $(u^j_k,Z^j_{k|k-1,u^j_k})$ a pseudo-update is performed using Eqn. (\ref{eq:r_update_legacy}) - (\ref{eq:p_update_meas}) to compute the (pseudo) posterior density $\hat{f}^j_k(X^j_k|Z^j_{k|k-1,u^j_{k}},u^j_{k})$ from where the number of targets is estimated. We consider the variance of this estimate as a measure of uncertainty which should be minimized. More specifically let $\hat{f}^j_k = \{(r^{ji}_{k}(u^j_{k},Z^j_{k|k-1,u^j_{k}}), p^{ji}_{k}(.,u^j_{k},Z^j_{k|k-1,u^j_{k}}))\}^{v^j_{k,u^j_k}}_{i=1}$ then:

\begin{align}
	\hat{n}^j_k(u^j_{k},Z^j_{k|k-1,u^j_{k}})  &= \sum_{i=1}^{v^j_{k,u^j_k}} r^{ji}_{k}(u^j_{k},Z^j_{k|k-1,u^j_{k}}) \\
	\sigma^j_k(u^j_{k},Z^j_{k|k-1,u^j_{k}}) &= \sum_{i=1}^{v^j_{k,u^j_k}} r^{ji}_{k}(u^j_{k},Z^j_{k|k-1,u^j_{k}})~ \times \\
		& \quad( 1 - r^{ji}_{k}(u^j_{k},Z^j_{k|k-1,u^j_{k}}) ) \notag 
\end{align}

\noindent where $\sigma^j_k$ denotes the variance of the number of targets estimate. In essence, by minimizing the cardinality variance $\sigma^j_k$ agent $j$ chooses the control action which results in the best estimate regarding the number of targets. This variance is maximized when $r^{ji}_{k}(u^j_{k},Z^j_{k|k-1,u^j_{k}}) = 0.5,\forall i$ thus we can define the normalized variance as $\tilde{\sigma}^j_k = 4\sigma^j_k/v^{j}_{k}$, where we have dropped the dependence on the control action and measurements for notational clarity. 

Using the normalized variance we define the single agent objective function to be minimized as:
\begin{equation}
	\xi^j_\text{track}(u^j_{k},Z^j_{k|k-1,u^j_{k}}) = ( \tilde{\sigma}^j_k - 1) \sqrt{\frac{\hat{n}^j_k}{\mathcal{V}_\text{cap}}} + 1
\end{equation}
where $\mathcal{V}_\text{cap}$ is a known design constant which specifies the maximum number of targets that our agent is able to track at a given time which is referred to as the agent tracking capacity in this paper. The above objective function is bounded in the closed interval $[0..1]$ and it reaches the minimum cost (i.e. best) when the agent reaches its maximum tracking capacity i.e. $\hat{n}^j_k = \mathcal{V}_\text{cap}$ with the maximum accuracy (i.e. $\tilde{\sigma}^j_k = 0$). On the other hand the agent receives the maximum cost when it tracks targets with the worst accuracy (i.e. $\tilde{\sigma}^j_k = 1$). It is not desirable to perform poorly in tracking mode thus the agent will try to find alternative controls for which the number of targets tracked is maximized and/or the variance is minimized. Additionally, an agent might even switch to search mode if this results in better collective search-and-track performance, rather than inaccurately tracking targets. It is worth noting that the agent will receive the maximum cost in the case where the number of targets being tracked is 0 i.e. $\hat{n}^j_k=0$. This behavior is by design since in this situation we wish to switch the agent to \textit{search} mode.  

\noindent Finally we define the multi-agent objective function for the tracking task as the average of their individual objective costs:
\begin{equation} \label{eq:trackScore_total1}
	 \frac{1}{|\alpha|} \underset{j \in \alpha}{\sum}  \xi^j_\text{track}(u^j_{k},Z^j_{k|k-1,u^j_{k}})
\end{equation}
where $\alpha \subseteq S$ denotes the set of agents in \textit{tracking} mode. As before we wish to minimize the objective function of Eqn. (\ref{eq:trackScore_total1}). In other words given a set of agents $\alpha \subseteq S$ we would like to drive the agents to the states which result in the minimization of Eqn. (\ref{eq:trackScore_total1}) or the maximization of the tracking accuracy.

\subsection{Multi-Agent Search Objective}
\begin{figure}
	\centering
	\includegraphics[width=\columnwidth]{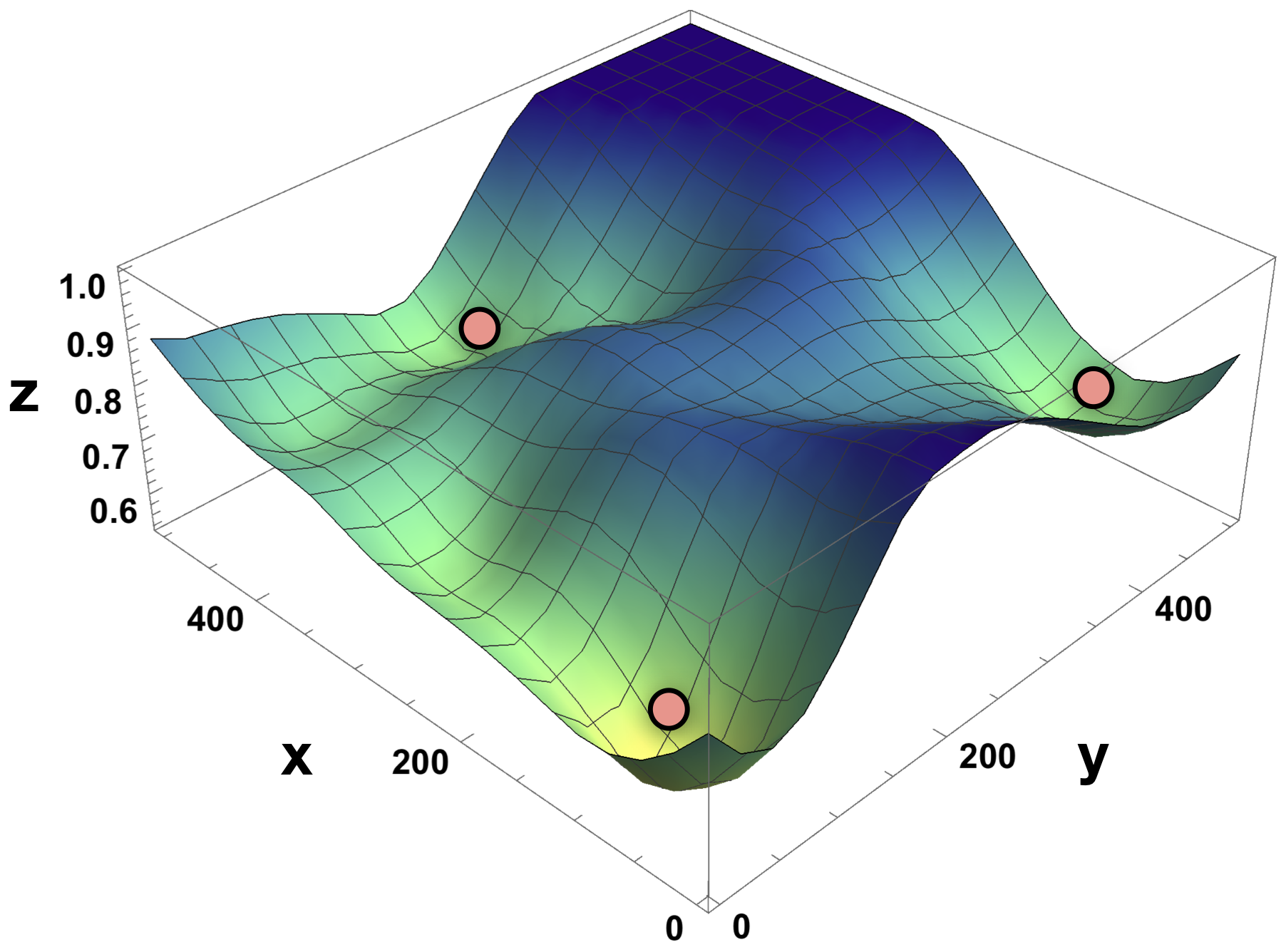}
	\caption{The figure shows an illustrative example of the search value (Z axis) i.e. Eqn. (\ref{eq:searchValueField}) over a surveillance region of size 500m by 500m for 3 agents. The position of agents are marked with red circles. In this figure dark areas indicate high search values i.e. in these regions targets are likely to not be detected  due to  the limited sensing range of the agents. Thus the agents have incentive to move towards these regions.}
	\label{fig:sv}
\end{figure}

In the previous sub-section we have defined the objective function for the task of tracking for single and multiple agents. In this section we are going to do the same for the task of searching. Let us define the \textit{search value} at time $k$ for specific location $p=(\textit{p}_x,\textit{p}_y) \in \mathcal{A}$ for the agent $j$ just after control action $u^j_{k} \in \mathbb{U}^j_{k}$ has been applied as:
\begin{equation}
	\xi^j(p, u^j_{k}) = 1-p_D(\tilde{\textit{p}},u^j_{k})
\end{equation}
where the function $p_D(\tilde{\textit{p}},u^j_k)$ is the agent's sensing model and $\tilde{\textit{p}} = [\textit{p}_x, 0, \textit{p}_y,0]$ constructs a location vector compatible with Eqn. (\ref{eq:sensing_model}). The idea of defining the search value as the complement of the probability of detection reflects the fact that locations with low probability of detection, should have increased value for searching. As a consequence distant locations with respect to the agent's location appear to have increased search value i.e. these areas are worth exploring in order to find new targets. Assuming pairwise detection independence between all agents and all targets, the search value at location $p$ when accounting for a set of agents $\alpha \subseteq S$ is given by: 

\begin{equation}\label{eq:searchValueField}
	 \xi^\alpha(p,\{u^j_k : j \in \alpha\}) = \underset{j \in \alpha}{\prod}~ \xi^j(p, u^j_{k})
\end{equation}
where $u^j_{k} \in \mathbb{U}^j_{k}$. This is shown in Fig. (\ref{fig:sv}). Finally, we define the total search value over the surveillance area, i.e. the multi-agent search objective function as:
\begin{equation}\label{eq:sv_total}
	\xi^\alpha_\text{search}(\{u^j_k : j \in \alpha\})=\frac{1}{\textit{A}}\int_\mathcal{A} \xi^\alpha(p,\{u^j_k : j \in \alpha\}) dp
\end{equation} 

\noindent where $\mathcal{A} \subset\mathbb{R}^2$  and \textit{A} is the total area of the 2D surveillance region. We should note here that the above coverage problem does not consider the current or future target locations whatsoever. However, it will be very interesting to investigate in the future a multi-agent coverage problem which takes into account locations with high probability of target appearance.

\section{Mathematical Programming Reformulation}
\label{sec:Optimization}

\renewcommand{\algorithmicrequire}{\textbf{Input:}}
\renewcommand{\algorithmicensure}{\textbf{Output:}}

To solve the proposed problem, we derive hereafter a mathematical program formulation ($\text{P}_1$) which takes into account the mode of operation and control actions in order to jointly optimize the search and track objective. Let $j\in\{1,\ldots,|S|\}$ represent all available agents and $i\in\{1,\ldots,|\mathbb{U}_k|\}$ the control actions for each agent. The following formulation considers all possible control actions i.e., $u_k^{j}\in\mathbb{U}_k^j$, as described in subsection \ref{ssec:AgentDynamics}. The binary variable $a_{ji}\in\{0,1\}$ identifies which control is taken by agent $j$ when in search mode, and binary variable $b_{ji}\in\{0,1\}$ indicates the control decision for agent $j$ when in tracking mode. Using this notation, $\text{P}_1$ can be transformed to the following binary non-linear program: 
\begin{align}
(\text{P}_2) \min & \>w \xi_\text{search}(a) +(1-w) \xi_\text{track}(b) \label{eq:objective}\\
\mathrm{s.t.} &\>\>\sum_{i=1}^{\mathbb{U}} a_{ji}+\sum_{i=1}^{\mathbb{U}} b_{ji}=1,\>\>\forall j\in\mathcal{|S|} \label{eq:mode}\\
&\>\> \norm{\lambda_1u_k^{ji}+\lambda_2u_k^{j'i}}> d_\text{min}\lambda_1\lambda_2 \label{eq:distanceconst}\\
&\lambda_1\in\{a_{ji},b_{ji}\},\lambda_2\in\{a_{j'i},b_{j'i}\}, \> j\neq j' \notag \\
&\>\> a_{ji}\in\{0,1\}, b_{ji}\in\{0,1\}, \label{eq:binconst}\\ 
& \forall j,j'\in\{1,\ldots,|S|\}, i\in\{1,\ldots,|\mathbb{U}_k|\} \notag
\end{align}

\noindent The objective function, as expressed in eq. (\ref{eq:objective}), is a convex sum (using scalar value $w$) of the search cost and the track cost achieved for a particular task assignment (i.e. either searching or tracking) and control decision (i.e. which control action to take next). The $w$ value is a user-defined parameter representing the level of emphasis given to the task of searching versus the task of tracking. The linear equality constraints in eq. (\ref{eq:mode}) ensure that each agent is only in the searching or tracking mode at any instance in time and only a single control decision in made in that time instance. Finally, eq. (\ref{eq:distanceconst}) ensures that all agents are a minimum distance $d_\text{min}$ apart at any point in time which is achieved by computing the distance for any control points (i.e., $u_{k}^{ji}$) multiplied by the decision made in any of the 2 modes of operation (searching or tracking) indicated by variables $a_{ji},\>\>b_{ji}$ using indicators $\lambda_1,\>\lambda_2$. Specifically, variables $\lambda_1,\>\lambda_2$ check every possible combination of $a_{ji}$ and $b_{ji}$ values against all other $a_{j'i}$ and $b_{j'i}$ values to ensure that agents in either search or track mode are not closer that a minimum distance $d_\text{min}$ from each other.

\begin{algorithm}
\caption{: Proposed decision and control algorithm}
\label{alg:GAalgorithm}
\begin{algorithmic}[1]
\REQUIRE $f_{k-1}^j, \>\>\forall j\in\mathcal{S}$
\STATE Calculate the predictive multi-object density $f_{k|k-1}^j$ using Eqn. (\ref{eq:RFS_prediction1})-(\ref{eq:RFS_prediction3}).
\STATE Estimate the number of targets from $f_{k|k-1}^j$ as 
$\hat{n}^j_{k|k-1} = \sum^{v^j_{k|k-1}}_{i=1} \vecbold{1}_{(r^{ji}_{k|k-1}>0.5)}$ and extract the corresponding multi-target state $\hat{X}^j_{k|k-1}$.
\REPEAT
\STATE iter++
\STATE Generate a candidate GA population of hypothesized control actions $u_k^{ji} \in \mathbb{U}^j_k, \forall j$ using Eqn. (\ref{eq:controlVectors}) with constraints (\ref{eq:mode}) - (\ref{eq:distanceconst}).
\STATE According to the generated hypothetical controls generate PIMS as $Z^j_{k|k-1} =Z^j_{k|k-1} \cup \{{\arg\max}_z ~ g_k(z|\hat{x}^j_{k|k-1},u_k^{ji}b^{ji})\}$ \quad $\forall ~~ \hat{x}^j_{k|k-1} \in \hat{X}^j_{k|k-1}$.
\STATE Perform the pseudo-update step using PIMS to calculate the posterior multi-object density $\hat{f}_k^j, \forall j$ using Eqn. (\ref{eq:RFS_update1})-(\ref{eq:p_update_meas}).
\STATE Let $\Lambda_\text{iter}=w \xi_\text{search}(a)+(1-w) \xi_\text{track}(b)$ according to (\ref{eq:objective})
\STATE Evaluate $\Lambda_\text{iter}$ and select the fittest solutions according to (\ref{eq:mode}) - (\ref{eq:binconst})
\UNTIL{$(\text{iter}<\text{iter}_{max})$} \AND  {$(\Lambda_\text{iter}-(\Lambda_{\text{iter}}-1)<\epsilon)$}
\RETURN optimal controls $\{{u^{j*}_k}   \},\forall j \in \mathcal{S}$
\STATE Apply optimal control action $u^{j*}_k$.
\STATE Receive the actual measurement set $Z^j_k$.
\STATE Compute posterior multi-object density $f^j_k$ using Eqn. (\ref{eq:RFS_update1})-(\ref{eq:p_update_meas}).
\STATE Estimate $\hat{n}^j_k$ and $\hat{X}^j_k$ from $f^j_k$ (as in Sec. \ref{ssec:trackObjective}) ,  $\forall j$.
\STATE Estimate total number of targets  $\hat{n}_k = \sum_j \hat{n}^j_k, \forall j$ and their states as $\hat{X}_k = \bigcup_j \hat{X}^j_k, \forall j$.
\end{algorithmic}
\end{algorithm}

Similarly to ($\text{P}_1$), the searching function can be expressed as follows:
\begin{equation}\label{eq:search_function}
\xi_\text{search}(a)=\frac{1}{\textit{A}}\int_{\mathcal{A}}\prod_{j=1}^{|S|}\prod_{i=1}^{|\mathbb{U}|} \left( 1-p^i_D(\tilde{\textit{p}},u_{k}^{ji}a_{ji}) \right) dp
\end{equation}
which is a product of the probabilities that no agent detects a target at a specific location $\tilde{\textit{p}}$ in the field. These probabilities are then integrated over the whole field $\mathcal{A}$ to compute the total search value of the particular combination of agents and specific control decisions. As shown in Eqn. (\ref{eq:search_function}) this search value is normalized by $\frac{1}{\textit{A}}$ where $\textit{A}$ is the maximum value attained when all probabilities equal to 0. The tracking accuracy can be expressed as follows:
\begin{equation} \label{eq:track_function}
\xi_\text{track}(b)=\frac{1}{\tau} \sum_{j=1}^{|S|}\sum_{i=1}^{|\mathbb{U}|} c_{ji}b_{ji}
\end{equation}
which is a sum of the costs $c_{ji}=\xi^j_\text{track}(u^{ji}_{k},Z^j_{k|k-1,u^{ji}_{k}})$ of an agent $j$ being in tracking mode taking control action $i$, when control decision $b_{ji}=1$. The tracking value can be normalized by $\tau = \sum_{j=1}^{|S|}\sum_{i=1}^{|\mathbb{U}|} b_{ji}$ which is the total number of agents in track mode.

Clearly, $\xi_\text{search}(a)$ jointly computes the search cost for a subset of agents being in search mode by the multiplication shown in Eqn. (\ref{eq:search_function}) that produce non-linear outputs. Coupled with the binary variables of the discrete control decisions results to an NP-hard optimization problem that is difficult to solve exactly in practice \cite{Cela1998}.
Nevertheless, for the particular structure of ($\text{P}_2$) and the binary encoding of the decision variables, a genetic algorithm (GA) can be applied to provide adequate solutions in practice. GA is a search-based technique that follows natural selection principles of solutions found to be relatively good \cite{Goldberg1989}. To do that, a population of candidate solutions are generated over the search space and evaluated using the objective function of the problem (fitness function). The fittest solutions are then recombined and mutated to generate new populations for the next iteration. The process terminates when specific conditions are met, including the objective function improvement over consecutive iterations or at a maximum number of iterations.


A genetic algorithm is used to solve ($\text{P}_2$) with the following key elements:
\begin{enumerate}
\item Binary decision variables $\{a_{ji}, b_{ji}\},\forall j \in S,\> i\in\{1,\ldots,|\mathbb{U}_k|\}$
\item The fitness function is reflected by the objective function in $(P2)$, i.e., Eqn. (\ref{eq:objective}), and so are the linear equality constrains in Eqn. (\ref{eq:mode}).
\item Control decisions that result to agents coming at a distance closer to $d_\text{min}$ are prune to satisfy constraint Eqn. (\ref{eq:distanceconst}).
\item Candidate solutions are generated around controls of the same mode of each agent.
\item The algorithm terminates after a maximum number of iterations has been reached or when the fitness function did not improve more than $\epsilon$.
\end{enumerate}

\noindent The complete algorithm of the proposed system is shown in Algorithm (\ref{alg:GAalgorithm}). The algorithm implements the recursion which is shown in Fig. (\ref{fig:sys_arch}) and discussed throughout this paper.

\section{Evaluation}
\label{sec:Evaluation}

\begin{figure*}
	\centering
    \begin{multicols}{2}
	   \includegraphics[width=0.9\columnwidth]{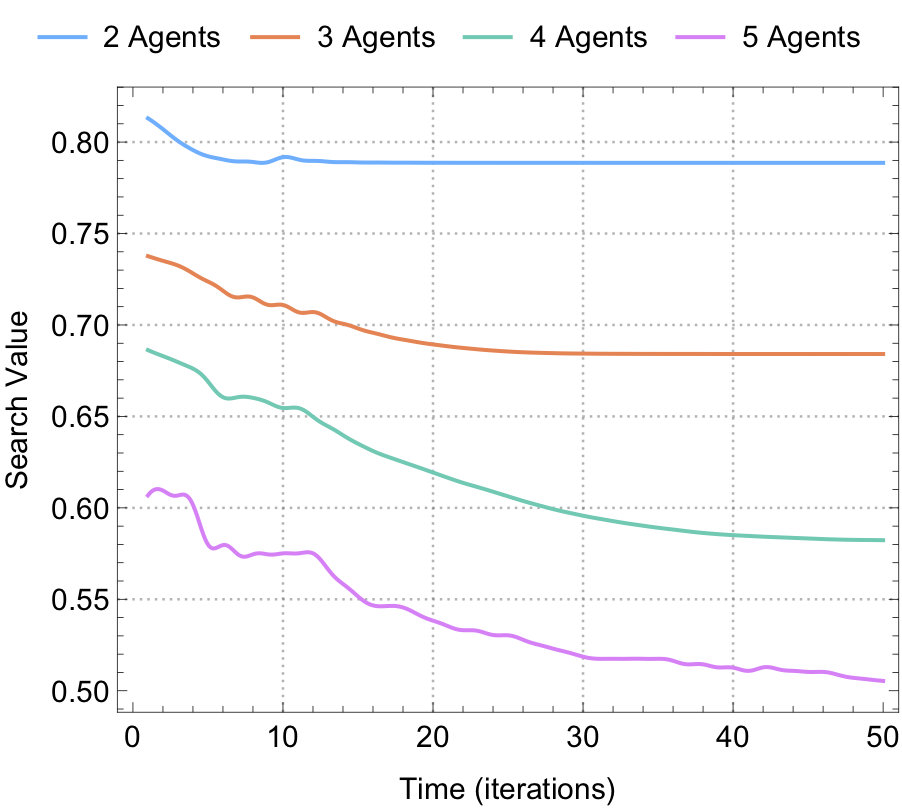}
	   \caption{The figure shows the evolution of the search cost for different number of agents inside a surveillance area of size 500m by 500m. As we can observe the search cost is decreasing over time which implies that the area coverage is increasing.}
	   \label{fig:sv_test}

	   \includegraphics[width=0.9\columnwidth]{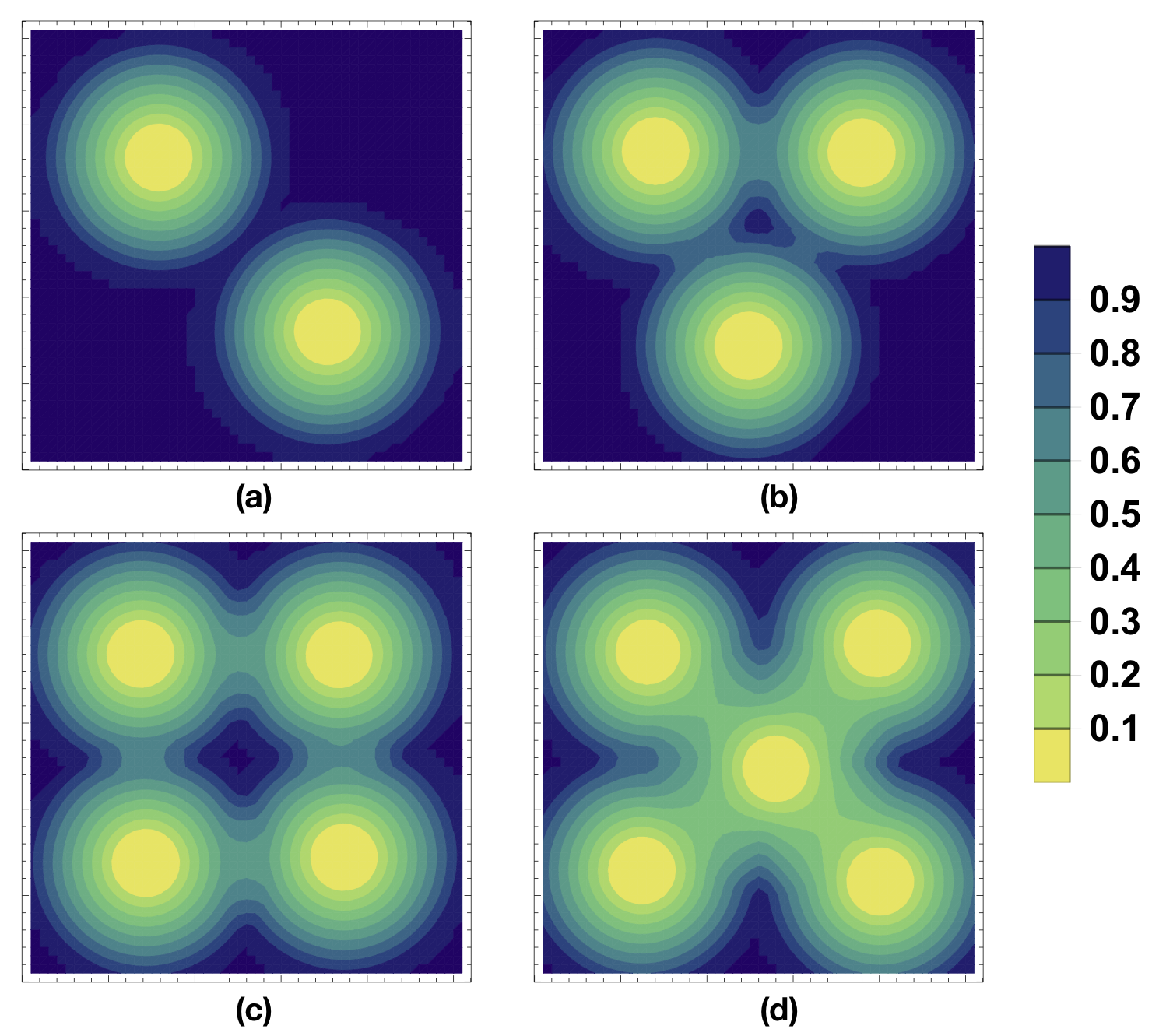}
	   \caption{The figure shows the search value over the whole surveillance area for different number of agents. As we can observe the agents take positions which result in the maximum area coverage (the figure shows the last time-step of a 50 time-step simulation experiment).}
	   \label{fig:sv_config}	   
	   
    \end{multicols}
\end{figure*}

\subsection{Implementation Details}

In order to evaluate the performance of the proposed approach we have conducted several numerical experiments involving various number of agents and targets for a variety of scenarios. In each experiment we compare the proposed approach with the ground-truth either qualitatively or quantitatively. In this section we include an extensive set of the conducted numerical simulations. 

More specifically, we assume that the targets maneuver in an area of 500m $\times$ 500m and that the single target state at time $k$ is described by $x_k = [p_{x},\dot{p}_x,p_y,\dot{p}_y]_k^\top$ where $(p_x,p_y)$ give the position of the target in Cartesian coordinates and $(\dot{p}_x,\dot{p}_y)$ are the velocities of the target in the $x$ and $y$ direction respectively. For the target dynamics in Eqn. (\ref{eq:single_dynamics}), we consider that each target is moving according to the near constant velocity model with the process noise being Gaussian. Thus the single target transitional density is given by $\pi_{k|k-1}(x_k|x_{k-1}) = \mathcal{N}(x_k;Fx_{k-1},Q)$ where:
\begin{equation*}
	F = \begin{bmatrix} 
        1 & T & 0 & 0\\
        0 & 1 & 0 & 0\\
        0 & 0 & 1 & T\\
        0 & 0 & 0 & 1\\
       \end{bmatrix}, ~
    Q = \begin{bmatrix} 
        T/3 & T/2 & 0 & 0\\
        T/2 & T & 0 & 0\\
        0 & 0 & T/3 & T/2\\
        0 & 0 & T/2 & T\\
       \end{bmatrix}
\end{equation*}
with sampling interval $T=1$s. Note that we have used a linear motion model in our evaluation for the target dynamics and so Eqn. (\ref{eq:single_dynamics}) becomes $x_k = F x_{k-1} + \text{v}_{k}$. The target survival probability from time $k-1$ to time $k$ is constant and does not depend on the target's state   i.e. $p_{S}(x_{k-1})=0.99$. 

\noindent Once the agent detects a target it receives bearing and range measurements thus the measurement model of  Eqn. (\ref{eq:meas_model}) is given by:
\begin{equation*}
	h_k(x_k,s_k) = \left[ \text{arctan}\left(\frac{s_y-p_y}{s_x - p_x}\right),~ \norm{s_k -Hx_k}_2\right]
\end{equation*}
where $H = \begin{bmatrix} 1 &0 &0 &0\\0 &0 &1 &0 \end{bmatrix}$. The single target likelihood function is then given by $g_k(z_k|x_k,s_k) = \mathcal{N}(z_k;h_k(x_k,s_k),\Sigma^\top \Sigma)$ and sigma is defined as $\Sigma = \text{diag}(\sigma_\phi,\sigma_\zeta)$. The standard deviations $(\sigma_\phi,\sigma_\zeta)$ are range dependent and given by:
\begin{align*}
    \sigma_\phi &= \phi_0 + \beta_\phi \norm{s_k-Hx_k}_2 \\
	\sigma_\zeta &= \zeta_0 + \beta_\zeta \norm{s_k-Hx_k}_2^2 
\end{align*}
with $\phi_0 = 2\pi/180$ rad, $\beta_\phi=10^{-5}~\text{rad}/\text{m}$, $\zeta_0 = 1$ m, and $\beta_\zeta = 5 \times 10^{-5}~\text{m}^{-1}$.
False alarm measurements (i.e. clutter) are generated with a Poisson rate $\lambda_k = 10$ uniformly distributed over the measurement space. The agent's sensing model parameters take the following values $p_D^{\text{max}} = 0.99$, $\eta = 23 \times 10^{-4}$ and $R_0=30$m.
The agent's dynamical model is shown in Fig. \ref{fig:controls} where the radial displacement $\Delta_R=5$m, $N_R=2$ and $N_\theta = 8$ which gives a total of 17 control actions, including the initial position of the agent. The tuning parameter $w$ which controls the weight given to searching and tracking is set to $0.5$ unless otherwise specified. Finally, the parameter $\mathcal{V}_\text{cap}$ is set depending on the simulation scenario to the maximum number of targets that we wish a single agent to be able to track. The genetic algorithm was implemented using the \textit{ga} Matlab function. We should note here that any optimization/meta-heuristic method which can handle the binary program of Eqn. (\ref{eq:objective}) with integer constraints, can be used in place of the genetic algorithm. However the genetic algorithm fits best to the structure of our problem. Finally, in order to handle the non-linear measurement model we have implemented a Sequential Monte Carlo (SMC) version \cite{Vo2009,BNVO2005} of the multi-Bernoulli filter. When the target dynamics and the measurement model are linear and the probability of detecting targets is state independent the Gaussian version of the multi-Bernoulli filter provides a computationally more efficient solution \cite{Vo2009}. Table \ref{tb:table1} shows a list of variables and common values used in the experimental evaluation of the proposed approach.
\begin{table}[!t]
\renewcommand{\arraystretch}{1.3}
\caption{List of variables used for simulations}
\label{tb:table1}
\begin{center}
	\begin{tabular}{| l | p{2.2cm} | p{3.5cm} |}
		\hline
		\textbf{Symbol} & \textbf{Description} & \textbf{Values, [Units]}\\
		\hline \hline
		$\mathcal{A}$ & Surveillance area & 500 $\times$ 500, [m]\\ \hline
		$x_k$ & Target state &  $[p_{x},\dot{p}_x,p_y,\dot{p}_y]_k^\top$ , [m, m$/$s, m, m$/$s]\\ \hline
		$T$ & Sampling interval & 1, [s] \\ \hline
		$p_S$ & Target probability of survival  & 0.99 \\ \hline
		$\sigma_\phi$ & Bearing meas. noise (std) & varies, [rad] \\ \hline
		$\sigma_\zeta$ & Range meas. noise (std) & varies, [m]\\ \hline
		$\lambda_k$ & False alarms rate & 10 [meas. per time-step] \\ \hline
		$p_D^\text{max}$ & Max prob. of detection & 0.99\\	\hline
		$\eta$ & Sensing model parameter & $23 \times 10^{-4}$\\	\hline
		$R_0$ & Sensing model parameter & 30 [m]\\	\hline
		$\Delta_R, N_R, N_\theta $ & Agent motion model params. & 5 [m], 2, 8\\	\hline
        GACT  & GA Constraint Tolerance & $10^{-6}$ \\ 	\hline
        GAFT ($\epsilon$) & GA Function Tolerance  & $10^{-4}$ \\ 	\hline
        GAPS  & GA Population Size & $400$ \\ 	\hline
        GAMG  & GA Max Generations & $150|\mathbb{U}||S|$ \\ 	\hline
        $w$ & Eqn. (\ref{eq:objective}) mode weight & 0.5 \\	
		\hline
	\end{tabular}
\end{center}
\end{table}

\color{black}


\subsection{Results}

The first set of experiments is conducted in order to investigate the proposed system on searching. More specifically the desired behavior is for the agents to jointly find the best configuration i.e. select the appropriate controls such that the total search value of the surveillance area is minimized or the area coverage is maximized. Our experimental setup is as follows. For a given number of agents we have conducted 50 Monte-Carlo (MC) trials where the agents are spawned at random locations (uniformly distributed) inside the surveillance area. Then we let our system to run for 50 time-steps and we monitor the total search value given by Eqn. (\ref{eq:sv_total}). Our intuition is that the total search value should decrease over time since the agents would try to move to locations where the overlap of their sensing ranges, is minimized. In other words, we would like to have the agents spread as much as possible inside the surveillance region to increase coverage. Figure \ref{fig:sv_test} shows the average total search value for each time-step of this experiment for 2, 3, 4, and 5 agents. As we can see from this figure, the total search value is decreasing over time, which indicates that the agents are trying to cover as much area as possible. It is also clear that as the number of deployed agents inside the surveillance region increases, the covered area increases as well.

In order to gain more insights into the behavior of the agents during searching, we have being monitoring the control actions assigned to them. At the end of the experiment, the final locations of the agents for the 4 cases are shown in Fig. \ref{fig:sv_config}. Interestingly, we can see that the agents take the appropriate formation which results in maximum coverage.
For instance from Fig. \ref{fig:sv_config}a, we see that when the number of agents is 2 the optimal configuration would be to take the diagonal of the surveillance area. On the other hand when we have 3 and 4 agents, the optimal configuration forms a triangle and a square respectively, whereas in the case of 5 agents, the last agent moves into the middle of the area to fill the gap. 
The previous experiments showed that the solution to the problem of Eqn. (\ref{eq:objective}) resulted in optimal coverage. In other words, depending on the number of agents, the obtained solution is a configuration of agent positions which maximizes the area coverage or minimizes the total search value. 

\begin{figure}
	\centering
	\includegraphics[width=0.9\columnwidth]{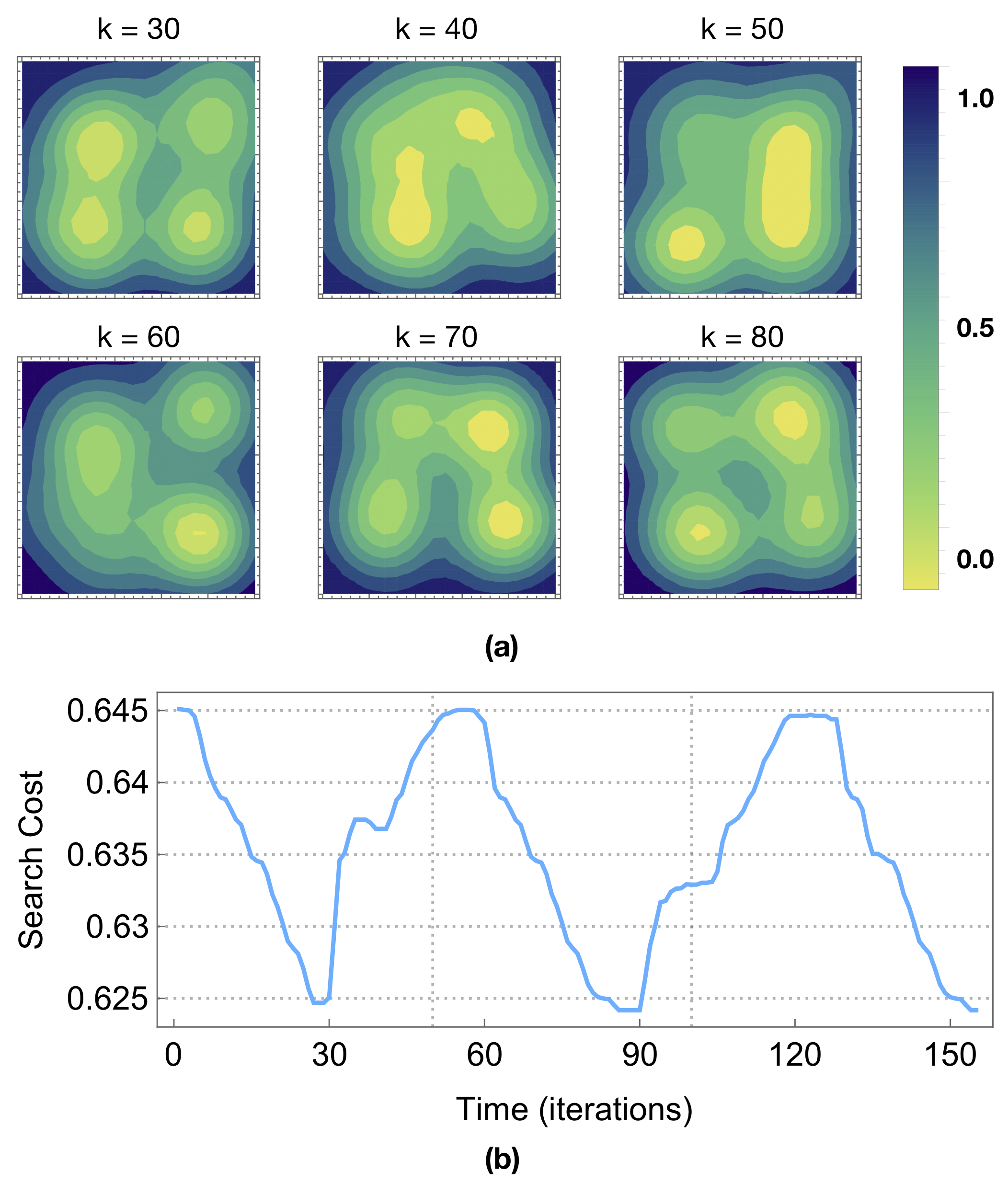}
	\caption{Memory-based searching: (a) The figure shows the evolution of the search value over the whole surveillance area for various time-steps of a 180 time-step experiment. As we can observe the agents are continuously moving inside the surveillance area to cover unexplored areas. (b) the corresponding search cost over same experiment.}
	\label{fig:sv_config_mem}
\end{figure}
\begin{figure*}
	\centering
	\includegraphics[width=1\textwidth]{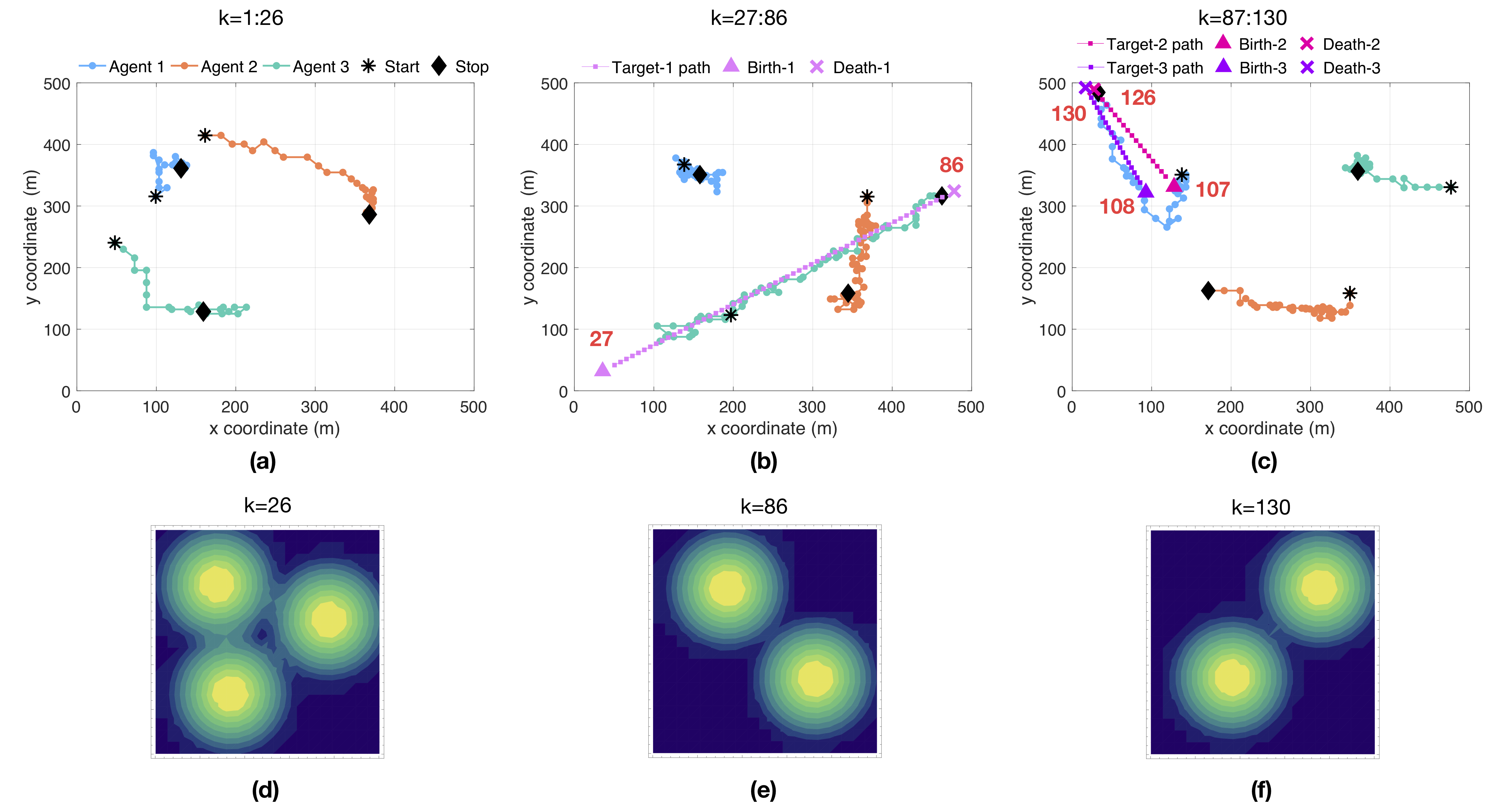}
	\caption{(a) - (c): The figure shows the maneuvers of 3 agents for the task of joint search and track in a simulated scenario during 3 time periods. Agent start and stop positions are marked with $\star$ and $\Diamond$, respectively. Target birth and death positions are marked with $\bigtriangleup$ and $\times$ respectively and their corresponding birth/death times are shown in red. (d) - (f): The figure shows the search value over the surveillance area for 3 time-steps i.e. $k=26, 86~ \text{and}~ 130$. As we can observe the agents in \textit{search} mode go to locations which result in the maximum area coverage.}
	\label{fig:eval_sys1}
\end{figure*}

\begin{figure}
	\centering
	\includegraphics[width=1\columnwidth]{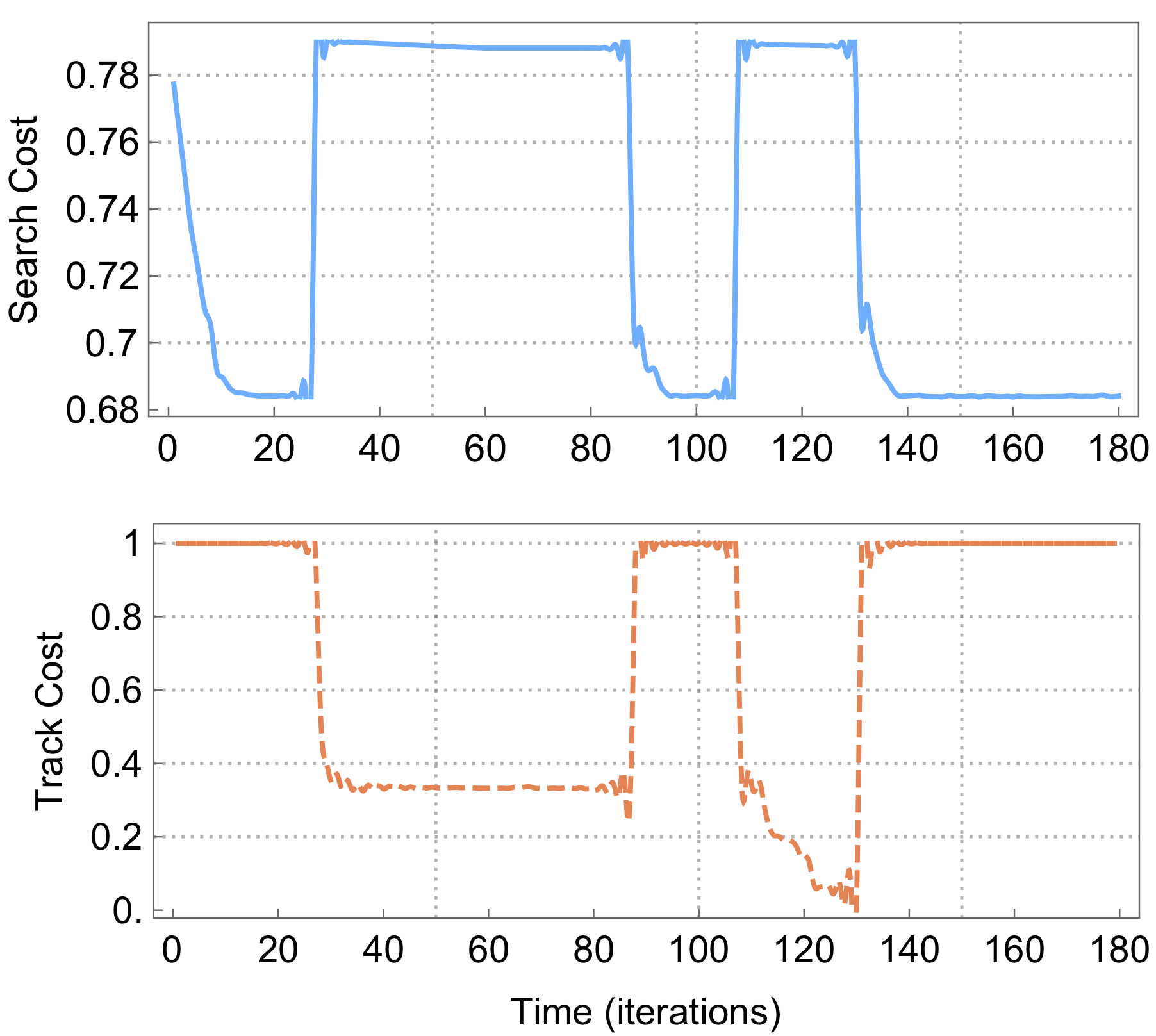}
	\caption{The figure shows the values of the search and track objective functions for a simulated 180 time-step scenario. At each time-step, the proposed system decides based on the cost of search and track which mode to assign to each agent so that the coupled objective is minimized.}
	\label{fig:searchTrack_Cost}
\end{figure}

We should note here that once the agents obtain a configuration which provides the optimal coverage (Fig. \ref{fig:sv_config}), they lock in that configuration (i.e. there is no movement) until something is changed (e.g. a target appears). This is quite reasonable since the search function of Eqn. (\ref{eq:search_function}) does not provide incentive for any movements once the optimal coverage is obtained. However, in certain situations it would be desirable to have the agents move to areas that have not been visited for some time and temporarily take a configuration which does not provide the optimal coverage. 

In order to achieve this we extend our proposed approach to include the notion of memory as follows: The total search value of Eqn. (\ref{eq:search_function}) now becomes the instantaneous total search value at time $k$. The new total search value is calculated as the weighted moving average of the $\kappa$ previous instantaneous total search values where $\kappa$ is the memory of our system. Figure \ref{fig:sv_config_mem} shows the memory-based searching behavior of our system for the case of 4 agents and $\kappa = 30$. More specifically, the figure shows the behavior of 4 agents in memory-based searching for 150 time-steps. In comparison with Fig. \ref{fig:sv_config}c we observe that the agents in this experiment are continuously moving in the surveillance area as shown in Fig. \ref{fig:sv_config_mem}a. As before the 4 agents at time $k=30$ take a square formation which gives the optimal coverage. However, since the system now takes into account the 30 previous time-steps, at time $k=40, 50,... $ we observe that a different formation has been taken. This is because the system \textit{remembers} what areas have been visited and pushes the agents to new locations. At $k=80$ the agents move again to form a square formation as shown in the figure. Fig. \ref{fig:sv_config_mem}b shows the evolution of the total memory-based search cost for this experiment. As we can observe there is an evident periodicity in this graph which is due to the memory effect.

Next, in order to verify the effectiveness of the proposed approach on the task of joint searching and tracking, we have conducted several experiments with different numbers of targets and agents. In this section, we will give a representative example with 3 agents and 3 targets. This scenario is depicted in Fig. \ref{fig:eval_sys1} during 3 different time-windows $k=1..26$, $k=27..86$ and $k=87..130$. 
More specifically, 3 agents enter the surveillance area of size $500\text{m} \times 500\text{m}$ at the locations marked with asterisks in Fig. \ref{fig:eval_sys1}a with coordinates $[100, 315]$,$[160, 415]$ and $[48, 240]$ for agents 1, 2 and 3, respectively. This experiment lasts 180 time-steps during which we have 3 target births and 3 target deaths. The ground-truth target birth/death times are $k=27/86$, $k=107/126$ and $k=108/30$ for targets 1, 2 and 3, respectively. The target birth locations are $[35, 31]$, $[128, 331]$ and $[92, 321]$ for targets 1, 2 and 3 respectively and their corresponding death locations are $[478, 324]$, $[27, 489]$ and $[16, 492]$. In Fig. \ref{fig:eval_sys1} the target birth locations are indicated by triangles, the target death locations are indicated by crosses and their birth/death times are shown in red. The trajectories of the agents and the targets during this experiment are shown in Fig. \ref{fig:eval_sys1}a -  \ref{fig:eval_sys1}c for periods $k=1..26$, $k=27..86$ and $k=87..130$ respectively. 

The agents enter the surveillance area in search mode at $k=1$ and at each time-step the proposed controller decides which mode (i.e. \textit{search} or \textit{track}) to assign to each agent. As we have already discussed each agent maintains a probability distribution regarding the number of targets and their location. The controller monitors the cost of tracking and the cost of searching and decides what are the optimal controls for all agents which minimize the objective function of Eqn. (\ref{eq:objective}). To summarize the cost of tracking takes into account whether or not targets exist in the area, the uncertainty of the estimated number of targets in the area and finally how many and how well targets are being tracked by each agent. On the other hand, the cost of searching takes into account how well the area is covered by the agents. 

Figure \ref{fig:eval_sys1}a shows the trajectories of all agents during period $k=1..26$. During this period no targets exist in the area and so the controller decides to assign all agents in search mode. The objective now is to increase coverage and so the agents move away from each other, trying to find the locations where the coverage is maximized. Figure \ref{fig:eval_sys1}d shows the search value for each location in the surveillance area for $k=26$, i.e. at the end of the period. As we can observe the agents take a triangle formation at the right locations which results in the maximum coverage. To understand better what is happening inside the controller, we have recorded the raw values of the search and track objective functions of Eqn. (\ref{eq:search_function}) and Eqn. (\ref{eq:track_function}) respectively for the whole experiment. This is shown in Fig. \ref{fig:searchTrack_Cost}. The first point to note here is that the search cost is decreasing during time-steps $k=1..26$, which indicates that the coverage is increasing. On the other hand, we can observe that the tracking cost during the same period is maximum, since the number of estimated targets in the area is 0. Thus the objective function of Eqn. (\ref{eq:objective}) is minimized when all agents are in search mode. 

Now, at time $k=27$ target 1 appears in the area as shown in Fig. \ref{fig:eval_sys1}b which is then being detected by agent 3 at the next time-step $k=28$. During time-steps $k=28..35$ agent 3 moves towards target 1 as shown in the figure and then for the rest of the period $k=36..86$, the same agent tracks target 1. It is also worth noting that during this whole period, agents 1 and 2 are in search mode, and are moving to locations which will result in optimal coverage. Figure \ref{fig:eval_sys1}e shows the search configuration of the two agents in \textit{search} mode and the resulting search values for the whole area. Figure \ref{fig:searchTrack_Cost} sheds some more light into this process. During time-steps $k=28..35$ we observe that the tracking cost is decreasing and afterwards during $k=36..86$ to be reaching a plateau. The tracking cost is decreasing during $k=36..86$ (i.e., the tracking accuracy is increasing) since the agent comes closer and closer to the target which allows the agent to perform better tracking. This is because the sensing and the measurement model depend on the distance of the agent to the targets. For distant targets, the uncertainty is higher thus the tracking accuracy lower. Once the target is locked in, the cost of tracking reaches a plateau. On the other hand, the search cost increases instantly since only 2 agents (out of 3) are left in search mode, and thus the coverage has been decreased. There is however a small improvement in the search cost during this period as the agents take a formation which results in better coverage. 

Target 1 dies at time-step $k=86$, which drops the number of estimated targets, of agent 3, from 1 to 0 at $k=87$. This increases the tracking cost to the maximum since no targets exist inside the surveillance area and as a consequence the controller assigns all agents to \textit{search} mode. This is shown in Fig. \ref{fig:searchTrack_Cost} for $k=87..107$. At $k=107$ the agents take a formation similar to Fig. \ref{fig:eval_sys1}d. During period $k=107..130$ two targets appear as shown in Fig. \ref{fig:eval_sys1}c which are being detected and tracked by agent 1. The tracking cost in Fig. \ref{fig:searchTrack_Cost} for this period reaches a new minimum since now two targets are being tracked and so the value of tracking over searching is increased. Agents 2 and 3 detect no target and remain in search mode moving towards locations which result in maximum coverage as shown in Fig. \ref{fig:eval_sys1}f. Finally, during period $k=131..180$, no targets are being detected, thus the controller assigns all agents to search mode. The tracking cost reaches the maximum and the agents optimize the coverage as shown in Fig. \ref{fig:searchTrack_Cost}.

Furthermore, we investigate hereafter, the impact of the parameter $w$ of Eqn. (\ref{eq:objective}) on the search-and-track behavior of the system. We assume the presence of 3 agents in locations $[100, 200]$, $[250, 150]$ and $[300, 400]$ for agents 1, 2 and 3, respectively, and we vary $w$ between $[0 ... 1]$ observing how this affects the operating mode (i.e. search or track) of each agent. We investigate 2 scenarios i.e. in the first scenario a target is located in position  $[350, 50]$ and in the second scenario a target is located in position  $[260, 140]$.
\begin{figure}
	\centering
	\includegraphics[width=1.0\columnwidth]{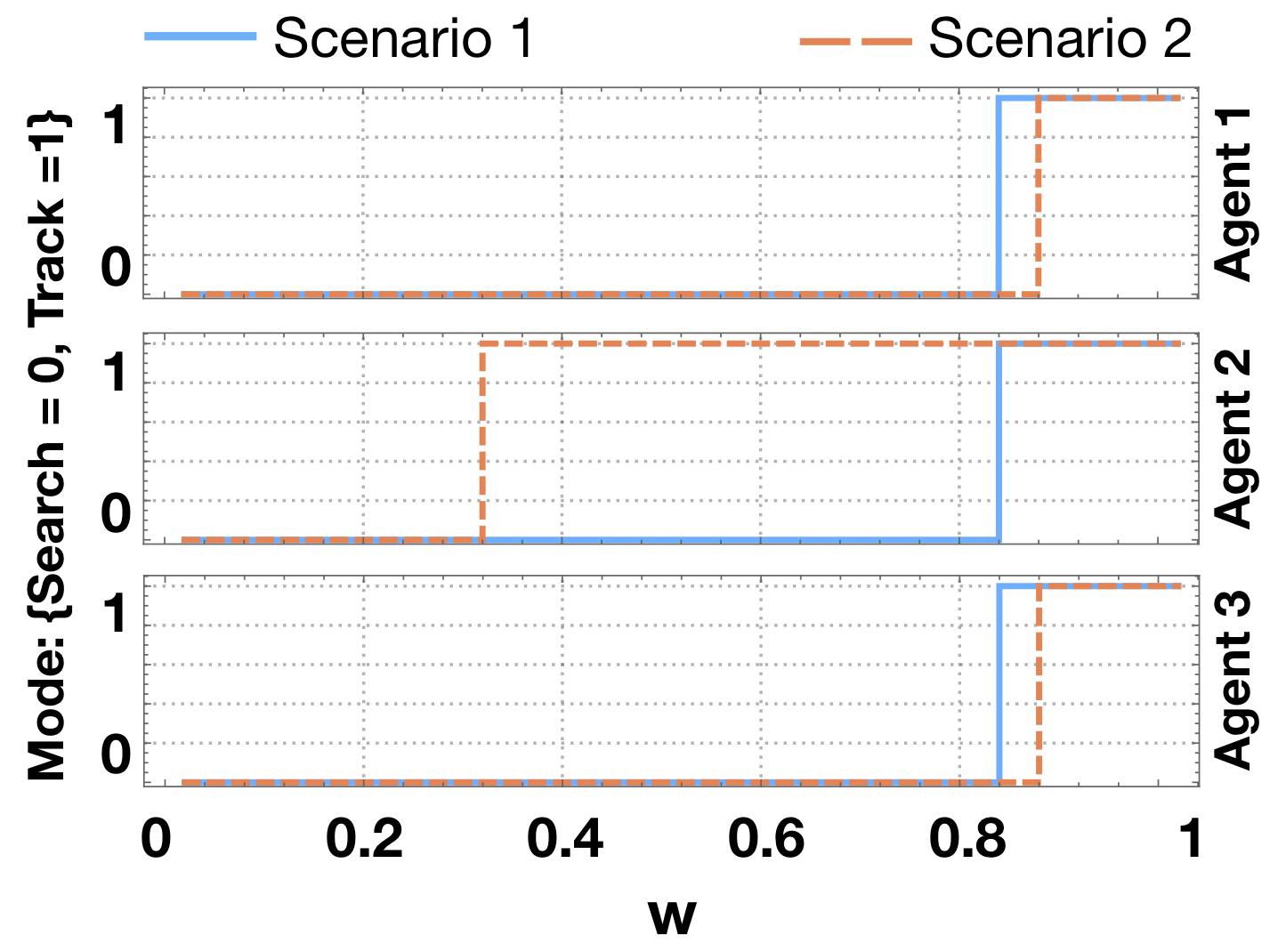}
	\caption{The figure illustrates the effect of the parameter $w$ on the search and track behavior of the system.}
	\label{fig:w}
\end{figure}
As we can observe from Fig. \ref{fig:w}, values of $w$ that are close to 0 force all agents to switch to search mode whereas values of $w$ very close to 1 have the opposite effect i.e. all agents are forced into track mode. This is a direct consequence of Eqn. (\ref{eq:objective}). Interestingly, in Scenario 1 the target located in $[350, 50]$ is far away from all the agents and thus for moderate values of $w$ close to $0.5$ the system chooses to have all agents in search mode. In particular, in this scenario the target is detected by agent 2 however, the normalized tracking variance $\tilde{\sigma}$ of this agent is as high as $0.9$ which results in very high tracking cost. This makes the system to prefer to have the particular agent in searching mode rather than in tracking mode for $w \le 0.83 $. In scenario 2 however, agent 2 is tracking the target located at $[260, 140]$ very accurately with $\tilde{\sigma}=0.05$. As a consequence the system assigns agent 2 in tracking mode for $w \ge 0.3$ as is shown in Fig. \ref{fig:w}. In this scenario however we can still force all agents in searching or tracking mode with the appropriate values for $w$. Dynamically adjusting $w$ depending on a given situation is something interesting which we will investigate in a future work. 

\begin{figure}
	\centering
	\includegraphics[width=1.0\columnwidth]{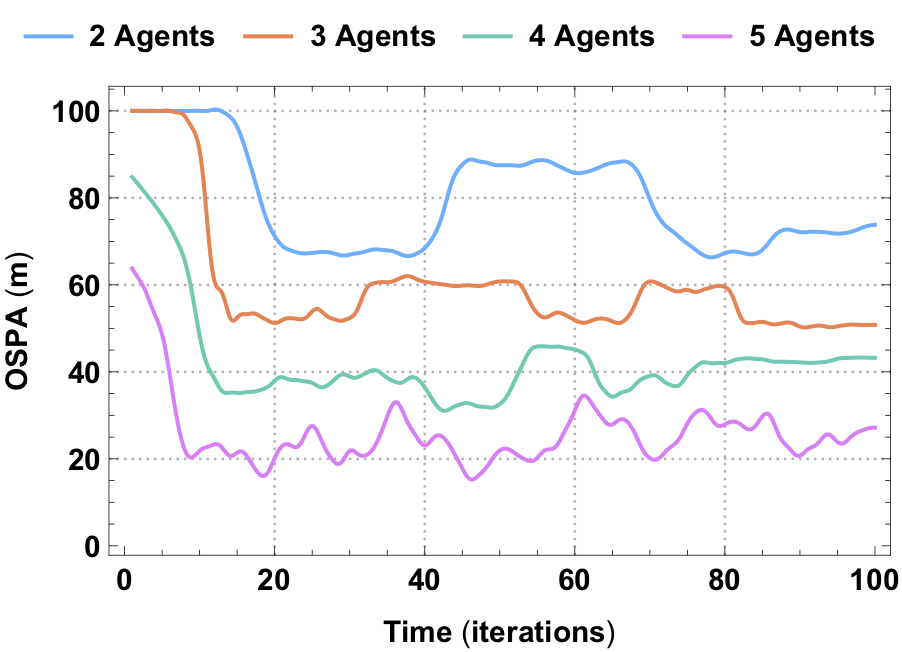}
	\caption{The figure shows the performance of the proposed system for searching and tracking 10 targets with 2, 3, 4 and 5 agents.}
	\label{fig:OSPA_agents}
\end{figure}

Moreover, we investigate the performance of the proposed approach with respect to the number of agents and the number of targets. In order to quantify the performance of the proposed approach we use the optimal sub-pattern assignment (OSPA) metric \cite{Schuhmacher2008} which is widely used to measure the accuracy of multi-object filters. Our first experiment is conducted as follows: We randomly generate 10 targets inside the surveillance area and we use OSPA to quantify the effectiveness of the proposed search-and-track approach with 2, 3, 4 and 5 agents. All agents are spawned from the center of the surveillance area at the start of each trial. This is done in order to simultaneously test the performance of search (i.e. coverage) and track. Intuitively, we would like to see the agents to start acquiring a configuration which results in maximum coverage and while doing so increasing their probability of finding and tracking the scattered targets in the surveillance region. Figure \ref{fig:OSPA_agents} shows the average OSPA error (with parameters $c=100$ and $p=2$) over 30 Monte Carlo trials for 100 time-steps for different number of agents. As we can see in all scenarios the tracking error starts high but drops significantly as time progresses. The agents start from the center of the surveillance area and jointly perform searching by covering as much surveillance region as possible. This is evident by the high OSPA error in the beginning of the experiment. In particular, we observe in the case of 2 agents that no targets are detected between time-steps 0 and 17. During that time the agents maximize coverage (i.e. see Fig. \ref{fig:sv_config}a). Around time-step 16 the OSPA error begins to drop which indicates that targets are being tracked. Between time-steps 20 to 40 the 2-agent system reaches its tracking capacity and achieves its best performance. Between time-steps 40-70 we can see that the error has been increased. We have observed that when this happens the agents are tracking single targets due to targets splits and target deaths which increases the tracking error (while elsewhere other targets might have been born which have not been detected by the already occupied agents). However, when no targets are being tracked, or the targets tracked are far away, the agents switch to searching in order to optimize coverage. This results in a decrease in the tracking error between time-steps 70 to 100. Similar behavior is also true for 3, 4 and 5 agents. As the number of agents increases the time to detect targets decreases as is shown in the figure. Finally, we observe that the tracking capacity of the system increases with the number of the agents, as a consequence the tracking error decreases as the number of agents increases.

In the next experiment we investigate how the number of targets affect the performance of our system. More specifically, we have fixed the number of agents to 2 and we varied the number of targets tracked by each agent. 
\begin{figure}
	\centering
	\includegraphics[width=1.0\columnwidth]{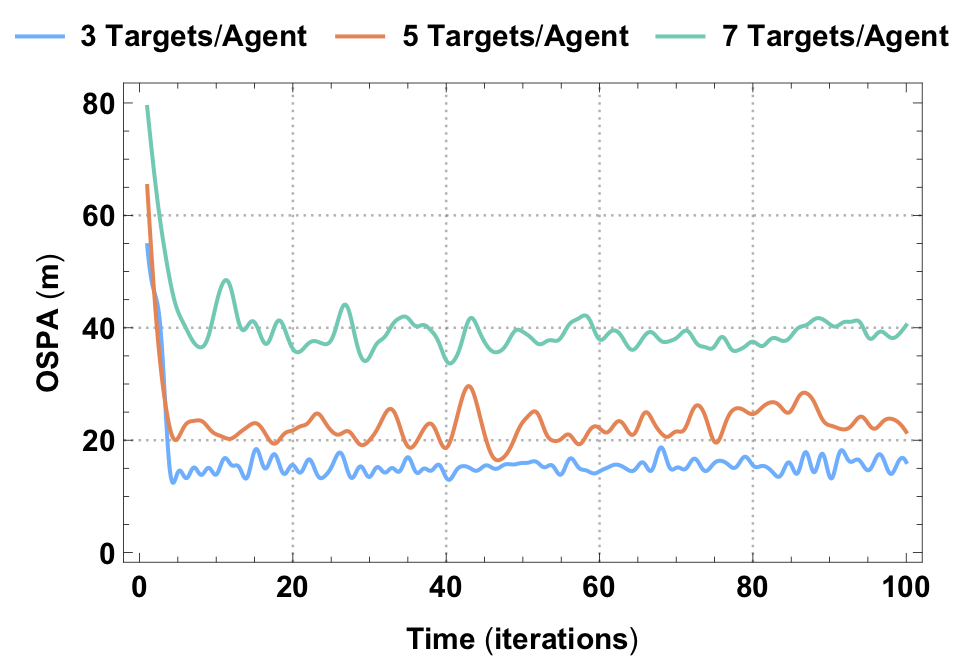}
	\caption{The figure shows the OSPA error when 2 agents track 3, 5 and 7 targets each.}
	\label{fig:OSPA_targets}
\end{figure}
Figure \ref{fig:OSPA_targets} shows the OSPA error when 2 agents track 3, 5 and 7 targets each for 100 time-steps. We should note here that in this experiment, the targets per agent (i.e. 3, 5 and 7) are initialized inside the corresponding agent's sensing range. Additionally the targets per agent move towards the same destination and without target splits occurring for the duration of the experiment. This is done here in order to investigate the ability of an agent to track 3, 5 and 7 targets. The figure shows the average error over 30 Monte Carlo trials. As we can observe, the tracking accuracy drops as the number of targets per agent increases. However, this experiment shows that the proposed approach is able to keep track multiple targets per agent. In particular, 3 and 5 targets per agent can be tracked with relatively high accuracy in this scenario. 

\begin{figure}
	\centering
	\includegraphics[width=1.0\columnwidth]{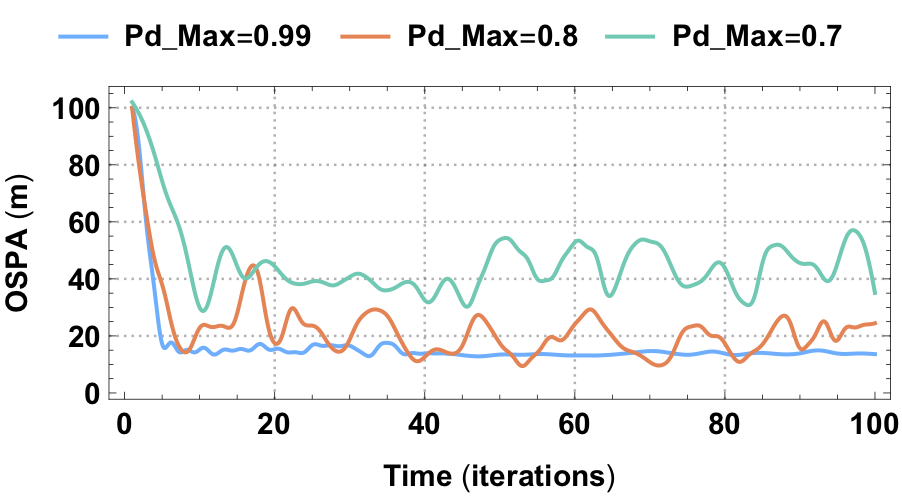}
	\caption{The figure shows the OSPA error when 3 agents track 3 targets with varying $p^{\max}_D$ values.}
	\label{fig:OSPA_pd}
\end{figure}

Up to this point, all of our experiments were based on the sensing model of Eqn. (\ref{eq:sensing_model}). In this model the parameter $p^{\max}_D$ determines the maximum detection probability of an agent with its value set to $0.99$. In order to investigate the impact of this parameter to the tracking accuracy of our system, 30 Monte-Carlo trials have been conducted of length 100 iterations, for 3 agents and 3 targets with varying values of $p^{\max}_D$. More specifically, the experimental set-up is as follows: for each Monte-Carlo run at time-step $k=0$, three agents are randomly generated inside the surveillance area. Then for each agent at $k=0$, a single target is spawned inside its sensing range. We leave the system to run for 100 time-steps and we measure the OSPA error. Figure \ref{fig:OSPA_pd} shows the average OSPA error over 30 trials for $p^{\max}_D = 0.99$,  $p^{\max}_D = 0.8$ and  $p^{\max}_D = 0.7$. As we can observe the best accuracy is achieved with a $p^{\max}_D$ value of $0.99$. On the other hand for $p^{\max}_D=0.8$ and $p^{\max}_D=0.7$ we observe significant OSPA oscillations which occur due to target miss-detections. More specifically, we have observed that a target miss-detection especially in the area beyond the agent's primary radius $R_0$ (where the probability of detection drops below $p^{\max}_D$) results in a) increased positioning error and b) the agent to take incorrect control action and/or switch mode (from tracking to searching) in subsequent time-steps which causes cardinality errors. Finally, note that the multi-Bernoulli filter (i.e. Eqn. (\ref{eq:RFS_prediction2})-(\ref{eq:p_update_meas})) is best suited for situations where the signal to noise ratio is high i.e. high probability of detection and the clutter rate is low. These problems can be alleviated with the more accurate labeled multi-Bernoulli (LMB) filter \cite{Reuter2014lrfs}. The LMB filter however, is computationally more expensive. 

\begin{figure}
	\centering
	\includegraphics[width=1.0\columnwidth]{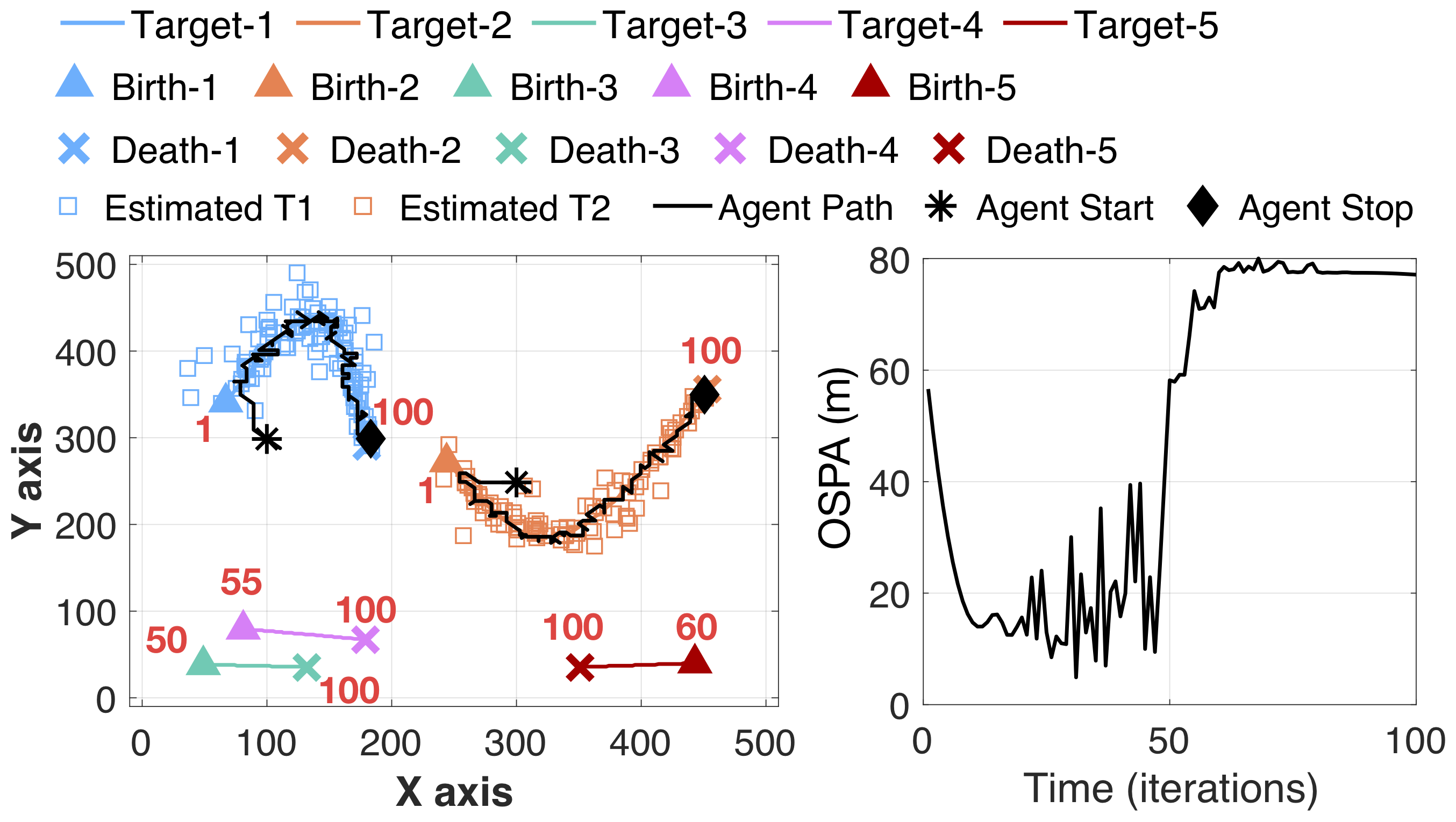}
	\caption{The figure shows a search-and-track scenario which illustrates the limitations of the proposed approach.}
	\label{fig:NL}
\end{figure}

Finally, we conclude the evaluation of the proposed approach by investigating some of its limitations. For this purpose we use the following setup: We generate 5 targets with initial positions $[67, 341]$, $[244, 272]$, $[49, 38]$, $[81, 79]$ and $[443, 40]$ for targets 1, 2, 3, 4 and 5, respectively. The birth/death times for the 5 targets are $k=1/100$, $k=1/100$, $k=50/100$, $k=55/100$ and $k=60/100$. Additionally at $k=0$ the position of two agents are $[100, 298]$ and  $[300, 248]$ for agents 1 and 2 respectively. This set-up is shown in Fig. \ref{fig:NL}. At $k=1$ both agents detect their nearby targets (target 1 and 2) and switch to track mode. The figure shows with blue and orange squares the estimated target positions. Initially the estimated target states exhibit large positioning errors however, as the figure shows, in subsequent time-steps the estimation error improves as the agents have locked-in the targets. The agent trajectories are shown with black lines. We should note here that in this scenario we use the linear target dynamics described in the beginning of this section to track the non-linear motion of target 1 and target 2. It is evident from Fig. \ref{fig:NL} that during time-steps $20$ to $50$ the OSPA error exhibits high oscillations which is due to the non-linear target behavior during that time-window. This is also shown by the high variance in the estimated target positions (blue and orange squares). At time-steps $50$, $55$ and $60$ targets 3, 4 and 5 appear in the surveillance region following the trajectories shown in the figure. These targets, however are not detected by any of the agents since both agents are occupied tracking target 1 and 2. As we can observe the OSPA error increases dramatically from $k=50$ onwards. This example shows the main limitation of the proposed system i.e. agents remain in a particular mode indefinitely. One way to approach this problem is by dynamically controlling the $w$ parameter and periodically forcing an agent to enter either the search or track mode depending on the situation. Another way will be to keep track of the visited areas in the surveillance region and store statistics regarding the time of visit, the number of targets detected at each area, etc. and then perform a priority-based search in order to maximize the number of detected targets. These ideas will be explored in future works.

\section{Conclusion and Future Work} \label{sec:Conclusion}
In this paper we have studied the problem of jointly optimized searching and tracking with multiple agents in stochastic environments. We have presented a novel unified probabilistic framework for this decision and control problem based on Bayesian multi-object stochastic filtering with random finite sets. We have defined suitable objective functions for the tasks of multi-agent searching and tracking and we have showed that the resulting non-linear binary program can be approximated adequately by a genetic algorithm. Finally, we have demonstrated the performance of the proposed decision and control algorithm through extensive numerical experimentation.  
Future work will look at simpler and computationally improved approximations of the non-linear binary program as well as alternative and simpler objective functions with equivalent behavior and performance. In addition, we are interested in investigating a rolling horizon (i.e. multiple-step look ahead) predictive control approach for this problem and analyzing its performance against the proposed single-step look ahead system. We are also interested in investigating this problem with heterogeneous agents i.e. agents which exhibit different types of sensing models.

\section*{Acknowledgments}
This work is supported by the European Union Civil Protection under grant agreement  No 783299  (SWIFTERS), by the European Union's Horizon 2020 research and innovation programme under grant agreement No 739551 (KIOS CoE) and from the Republic of Cyprus through the Directorate General for European Programmes, Coordination and Development.

\bibliographystyle{IEEEtran}
\bibliography{IEEEabrv,tmc18}  

\vspace*{10 mm}
\begin{IEEEbiography}[{\includegraphics[width=1in,height=1.25in,clip,keepaspectratio]{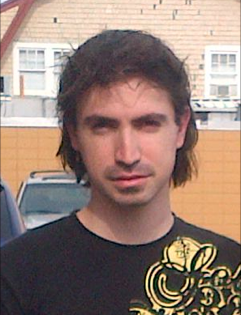}}]%
{Savvas Papaioannou} obtained his B.S. degree in Electronic and Computer Engineering from Technical University of Crete, his M.S degree in Electrical Engineering from Yale University and his Ph.D. degree in Computer Science from the University of Oxford. He is currently a Research Associate at the KIOS Research and Innovation Center of Excellence at the University of Cyprus. His research interests include intelligent systems, sensing platforms and architectures, sensor networks and sensor fusion, multi-target tracking and state estimation. He is a member of the IEEE and the ACM and also a reviewer for various journals.
\end{IEEEbiography}

\vspace*{20 mm}
\begin{IEEEbiography}[{\includegraphics[width=1in,height=1.25in,clip,keepaspectratio]{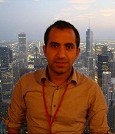}}]%
{Panayiotis Kolios} is a Research Associate at the KIOS Research and Innovation Center of Excellence at the University of Cyprus. He received his B.Eng and Ph.D degrees in Telecommunications Engineering from King's College London in 2008 and 2011, respectively. His interests focus on both basic and applied research on networked intelligent systems. Some examples of such systems include intelligent transportation systems, autonomous unmanned aerial systems, and the plethora of cyber-physical systems that arise within the Internet of Things. Particular emphasis is given to emergency response aspects in which faults and attacks could cause disruptions that need to be effectively handled. He is an active member of IEEE, contributing to a number of technical and professional activities within the Association. 
\end{IEEEbiography}

\vspace*{30 mm}
\begin{IEEEbiography}[{\includegraphics[width=1in,height=1.25in,clip,keepaspectratio]{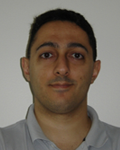}}]%
{Theocharis Theocharides} holds a Ph.D. degree in Computer Science and Engineering from the Pennsylvania State University (2006). He is currently an Associate Professor at the Department of Electrical and Computer Engineering, and the Director of Research at the KIOS CoE, both at the University of Cyprus. His research focuses on the broad area of intelligent embedded systems design, with emphasis on domain-specific architectures, evolvable and reconfigurable hardware, and design of reliable and low power embedded and application specific processors and circuits. He is a senior member of the IEEE and the IEEE Computer Society, an ACM Member, and a member of the HiPEAC Network of Excellence. Currently, he serves on the editorial boards of the IEEE Consumer Electronics Magazine, the ACM Journal on Emerging Technologies in Computing Systems, IEEE Design and Test magazine, and on several Organizational and Technical Program Committee boards of various IEEE/ACM Conferences.
\end{IEEEbiography}

\vspace*{-30 mm}
\begin{IEEEbiography}[{\includegraphics[width=1in,height=1.25in,clip,keepaspectratio]{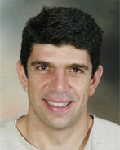}}]%
{Christos Panayiotou} has received a B.Sc. and a Ph.D. degree in Electrical and Computer Engineering from the University of Massachusetts at Amherst, in 1994 and 1999 respectively. He also received an MBA from the Isenberg School of Management, at the aforementioned university in 1999. Currently he is an Associate Professor at the Electrical and Computer Engineering Department at UCY and he serves as the Deputy Director of the KIOS CoE for which he is also a founding member. His research interests include distributed control systems, wireless, ad hoc and sensor networks, computer communication networks, quality of service (QoS) provisioning, optimization and control of discrete-event systems, resource allocation, simulation, transportation networks and manufacturing systems. He is a senior member of the IEEE and also a reviewer for various conferences and journals, and he has served in the organizing and program committees of various international conferences.
\end{IEEEbiography}

\vspace*{-30 mm}
\begin{IEEEbiography}[{\includegraphics[width=1in,height=1.25in,clip,keepaspectratio]{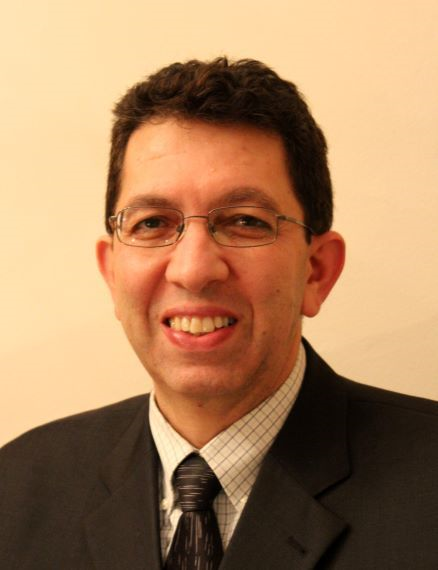}}]%
{Marios M. Polycarpou} is a Professor of Electrical and Computer Engineering and the Director of the KIOS Research and Innovation Center of Excellence at the University of Cyprus. He received the B.A degree in Computer Science and the B.Sc. in Electrical Engineering, both from Rice University, USA in 1987, and the M.S. and Ph.D. degrees in Electrical Engineering from the University of Southern California, in 1989 and 1992 respectively. His teaching and research interests are in intelligent systems and networks, adaptive and cooperative control systems, computational intelligence, fault diagnosis and distributed agents. Dr. Polycarpou has published more than 300 articles in refereed journals, edited books and refereed conference proceedings, and co-authored 7 books. He is also the holder of 6 patents.
Prof. Polycarpou is a Fellow of IEEE and IFAC. He is the recipient of the 2016 IEEE Neural Networks Pioneer Award. He received with his co-authors the 2014 Best Paper Award for the journal Building and Environment (Elsevier). Prof. Polycarpou served as the President of the IEEE Computational Intelligence Society (2012-2013), and as the Editor-in-Chief of the IEEE Transactions on Neural Networks and Learning Systems (2004-2010). He is currently the President of the European Control Association (EUCA). Prof. Polycarpou has participated in more than 60 research projects/grants, funded by several agencies and industry in Europe and the United States, including the prestigious European Research Council (ERC) Advanced Grant.
\end{IEEEbiography}

\end{document}